\documentclass[aps,twocolumn,showpacs,amsmath,amssymb,superscriptaddress]{revtex4-1}
\usepackage[utf8]{inputenc}

\usepackage{tikz}
\usepackage{graphicx}
\usepackage{amsfonts}
\usepackage{amsmath}
\usepackage{amssymb,bbold}
\usepackage{color}
\usepackage{bm}
\usepackage{bbm}
\usepackage{enumerate}
\usepackage{amsfonts, amsmath, amsthm, amssymb} 
\usepackage{mathtools}
\usepackage{array}
\usepackage[colorlinks,bookmarks=false,citecolor=blue,linkcolor=red,urlcolor=blue]{hyperref}

\usepackage[normalem]{ulem}

\begin{document}
\title{Long-Range Entangled-Plaquette States\\ for Critical and Frustrated Quantum Systems on a Lattice}

\author{J\'{e}r\^{o}me Thibaut}
\author{Tommaso Roscilde}
\author{Fabio Mezzacapo}
\affiliation{Univ Lyon, Ens de Lyon, Univ Claude Bernard, CNRS, Laboratoire de Physique, F-69342 Lyon, France}
\date{\today}	

\begin{abstract}											
We explore a variational Ansatz for lattice quantum systems -- named long-range entangled-plaquette state --  in which pairs of clusters of adjacent lattice sites are correlated at any distance. The explicit scale-free structure of correlations built in this wave function makes it fit to reproduce critical states with long-range entanglement. The use of complex weights in the Ansatz allows for an efficient optimization of non positive definite states in a fully variational fashion, namely without any additional bias (arising \emph{e.g.} from pre-imposed sign structures) beyond that imposed by the parametrization of the state coefficients. These two features render the Ansatz particularly appropriate for the study of quantum phase transitions in frustrated systems. Moreover,  the Ansatz can be systematically improved by increasing the long range plaquette size, as well as by the inclusion of even larger adjacent-site plaquettes. We validate our Ansatz  and its optimization procedure in the case of the XX and Heisenberg chain, and further apply it to the case of a simple, yet paradigmatic model of frustration, namely the  $J_1-J_2$ antiferromagnetic Heisenberg chain. For this model we provide clear evidence that our trial wave function faithfully describes both the short-range physics (particularly in terms of ground state energy)  and the long-range one expressed by the Luttinger exponent, and the central charge of the related conformal field theory, which govern the decay of correlations and the scaling of the entanglement entropy, respectively. Finally we successfully reproduce the incommensurate correlations developing in the system at strong frustration, as a result of the flexible representation of sign (phase) structures via complex weights.
	
 \end{abstract}


\maketitle

\section{Introduction} 

 Knowing the equilibrium state of quantum many-body systems is one of the central problems of modern theoretical physics. Similar to the classical case, this problem can be generally cast in the form of the evaluation of a statistical sum over a number of configurations growing exponentially with the number  of degrees of freedom, and addressed with a stochastic  approach, defining the general strategy of quantum Monte Carlo (QMC) techniques \cite{Sandvik2010, BeccaS2017}. In contrast to the classical case, however,  the statistical sum may have weights which are not positive definite (or may even be complex) leading to the so called sign (or phase) problem \cite{ChandrasekharanW1999}; the latter imposes the price of an exponentially large statistics (in the system size or in the inverse temperature) for reliable results to be obtained. Such a fundamental limitation currently impairs significantly our understanding of strongly correlated fermionic systems (from models of electrons in solids to models of elementary particles in relativistic quantum field theory), frustrated quantum magnetism, and bosonic quantum particles in gauge fields, to cite a few examples.
 
  In this context, a general alternative strategy is offered by the variational approach \cite{BeccaS2017}, mostly focusing on the ground state of the many-body problem of interest, and consisting in a chosen  parametrization (Ansatz) of the ground-state wave function in terms of a reduced  number of parameters (polynomial in the system size). In the following, we shall specialize our discussion to $S=1/2$  quantum spin models on a lattice, and to the corresponding bosonic or fermionic Hamiltonian that they can be mapped onto. Hence, the general form of the wave function reads:
  \begin{equation}
  |\Psi\rangle = \sum_{\bm \sigma} \psi(\bm \sigma) |\bm \sigma\rangle
  \end{equation}
  where, for a lattice of $N$ sites, $|\bm \sigma \rangle = |\sigma_1, ..., \sigma_N \rangle$ is the eigenvector of \emph{e.g.} the $S_i^z$ operator at each site $i$ ($\sigma_i = \pm 1/2$). Providing a variational wave function ultimately means choosing a suitable form of the wave function coefficients (or weights) $\psi(\bm \sigma) \rightarrow \psi(\bm \sigma;\bm C)$ depending on a set of adjustable parameters $\bm C = \{ C_l \}$, where the index $l$ is here used to enumerate the elements of $\bm C$.
 Over the last decades many variational Ans\"atze have been formulated to describe the ground-state physics of lattice spin models escaping the reach of unbiased QMC approaches. A few examples are matrix product states (MPS) \cite{Schollwoeck2005, Schollwoeck2011}, various tensor network states (TNS) \cite{Orus2014,Verstraeteetal2008}, resonating valence bond (RVB) states \cite{Liangetal1988, Beccaetal2010}, neural network quantum states (NNQS) \cite{CarleoT2017}, as well as entangled-plaquette states (EPS) \cite{Mezzacapoetal2009}, which are also known as correlated-product states (CPS) \cite{Changlanietal2009}, and constitute the focus of this work. The crucial aspect for the success of a variational Ansatz is its ability to reproduce the entanglement and correlations expected for the ground state of the Hamiltonian of interest. The entanglement content and correlation properties of MPS and different TNS have been extensively characterized \cite{Schollwoeck2011} and recognized as one of their main limitations, while other variational states are more flexible on the entanglement content \cite{GaoD2017}. An important bias, common to any variational form, is offered by the specific parametrization of  the wave function coefficients  $\psi(\bm \sigma)$, and in particular of their \emph{sign} structure, which generally becomes non-trivial when dealing with the ground state of a frustrated spin model \cite{BeccaS2017}. In the case of RVB states the sign structure of the coefficients is inherited from that of a fermionic state projected onto the spin Hilbert space to provide the Ansatz \cite{BeccaS2017}; only a few studies have dealt so far with the sign structure of NNQS, either imposing it a priori \cite{CarleoT2017, Chooetal2019} or modeling it using an auxiliary neural network \cite{Caietal2018}. 
 
  In this work we introduce the long-range entangled-plaquette states (LR-EPS) which offer a very flexible parametrization of highly entangled quantum spin wave functions. The major strength of EPS in general relies on the possibility of explicitly correlating different groups of lattice sites into plaquettes, expressing  $\psi(\bm \sigma)$ as a product of plaquette coefficients $C_p(\bm \sigma_p)$, where $p$ is the plaquette index and $\bm \sigma_p$ is the plaquette configuration. Designing the EPS Ansatz by considering overlapping (\emph{i.e.} entangled) plaquettes constitutes a fundamental aspect which will be discussed in the following section.  LR-EPS   are a natural generalization of the entangled-plaquette wave functions  based on plaquettes of adjacent sites (hereafter referred as A-EPS) mostly adopted in previous works on the subject \cite{Mezzacapoetal2009, MezzacapoC2010, Mezzacapo2011, Al-Assametal2011, MezzacapoB2012, Mezzacapo2012, Duricetal2014, Duricetal2016, Mezzacapoetal2016}. The idea of the  LR-EPS Ansatz is to consider plaquettes directly correlating clusters of adjacent lattice sites  at arbitrary distances. This results in a potentially \emph{scale-invariant} wave function which can accurately capture long-range entanglement and effectively describe critical ground states (associated with quantum critical points, gapless ordered phases or extended critical ones). Moreover the use of \emph{complex} plaquette coefficients $C_p(\bm \sigma_p)$ allows one to reproduce various sign/phase structures, turning the latter into a feature that can be fully optimized variationally. Hence, the LR-EPS Ansatz marries the relative simplicity of the EPS concept with a high degree of flexibility in coding the correlations and the sign structure of a quantum state. Moreover, as  any EPS wave function, it can be systematically improved by increasing the size of the (long-range) plaquettes, as well as by combining adjacent-site and long-range plaquettes. We first validate our Ansatz in the case of the $S=1/2$ XX chain. Remarkably the ground state of this model, along with that of other exactly solvable ones \cite{Sutherland-book}, is a LR-EPS, whose form our optimization algorithm is able to accurately reconstruct without any bias except for the constraint of translational invariance. This is proven by our variational results being able to reproduce the exact correlations and entanglement structure of the $S=1/2$ XX chain with excellent precision. 
We then move on to the frustrated $J_1-J_2$ chain, for which our Ansatz is shown to successfully reproduce the most challenging traits of the physics, namely the quantum phase transition from gapless spin liquid to valence-bond crystal; and the appearance of incommensurate correlations as the degree of frustration is increased. In particular we provide extensive results for the evolution of entanglement properties across the above cited transition. \\
The paper is structured as follows. Sec.~\ref{s.LREPS} briefly recalls the basics of the EPS wave function, presents in detail the form of the LR-EPS Ansatz, its parametrization of the sign structure, the optimization strategy, and the observables relevant to this study; Sec.~\ref{s.XX} shows a validation of the Ansatz in the case of the exactly solvable XX chain while Sec.~\ref{s.J1J2}, discusses our results for the $J_1-J_2$ chain; conclusions and perspectives are presented in Sec.~\ref{s.conclusions}. 
 \begin{figure}[t]
\includegraphics[width=0.9\columnwidth]{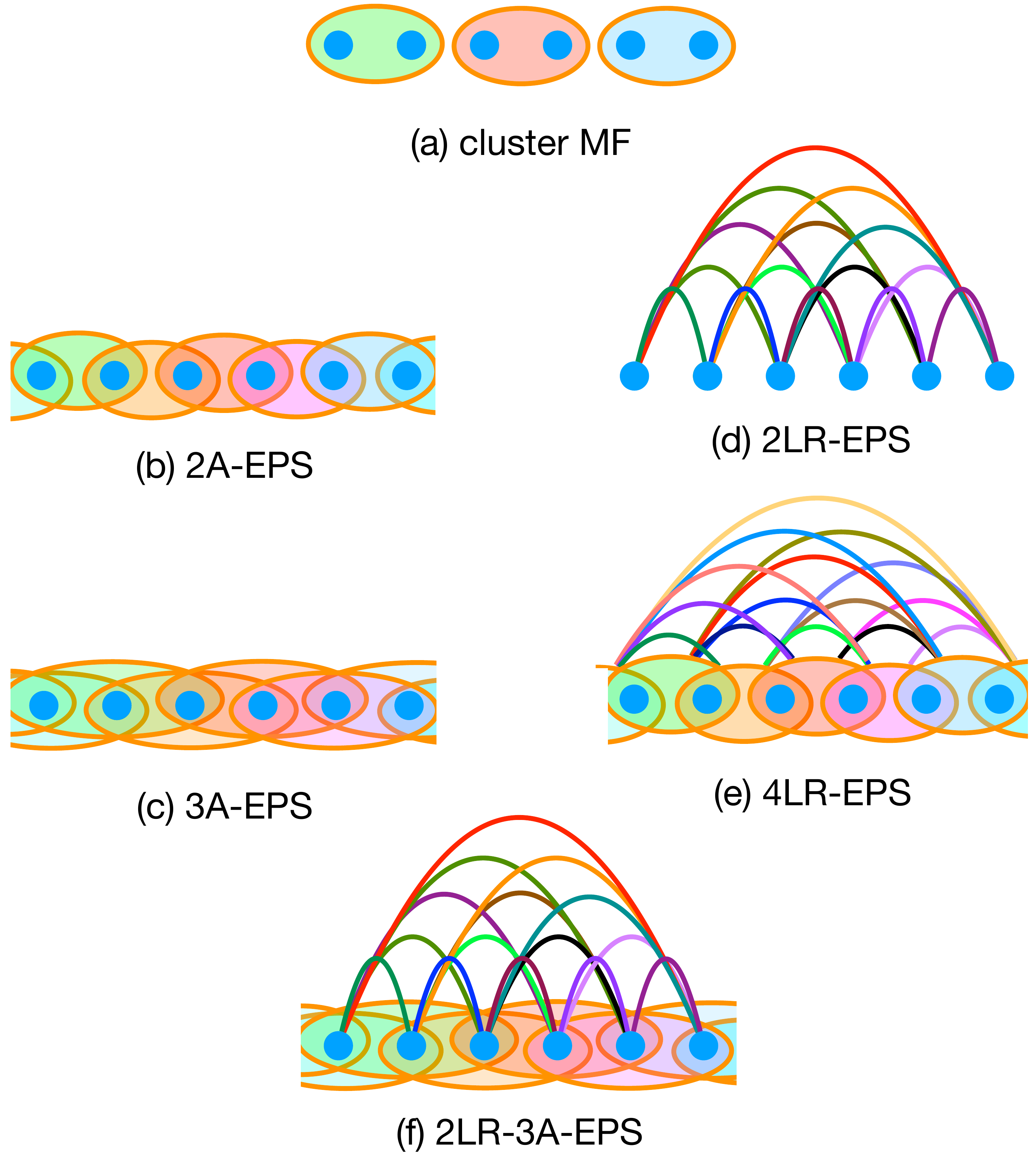}
\caption{Sketch of the variational Ans\"atze of interest to this work. Here the colorful ellipses indicate plaquettes made of adjacent sites; arcs indicate plaquettes composed of sites at arbitrary distance; and ellipses linked by arcs indicate plaquettes composed of two non-overlapping clusters, each made of adjacent sites. The various Ans\"atze represented here are: the cluster mean-field one (a); the adjacent-site EPS (A-EPS) with plaquettes of size $n=2$ (b), and $n=3$ (c); the long-range EPS (LR-EPS) with plaquettes of size $n=2$ (d) and $n=4$ (e); and the $n$LR-$n'$A-EPS Ansatz with $n=2$ and $n'=3$ built via the combination of LR and adjacent-site plaquettes of size $n$ and $n'$, respectively (f).} 
\label{Fig:1}
\end{figure} 
  
 \section{Long-range entangled-plaquette states: form, optimization and observables}
\label{s.LREPS}
 
 \subsection{Structure of the Ansatz}
 
  The EPS Ansatz amounts to parametrizing the wavefunction weights as 
  \begin{equation}
  \psi_{\rm EPS}(\bm \sigma) = \prod_p C_p(\bm \sigma_p)
  \label{e.EPS}
  \end{equation}
where $C_p(\bm \sigma_p)$ are coefficients associated with plaquettes indexed with $p$ and of size $n_p$, and $\bm \sigma_p = (\sigma_{1,p}, ..., \sigma_{n_p,p})$ is the plaquette configuration in the computational basis ($j,p$ is the index of the $j$-th site of the $p$-th plaquette). In general terms, the plaquettes may have a completely arbitrary geometry, although their size $n_p$ is necessarily limited to $\sim {\cal O}(10)$, due to the exponential growth of the parameter space with $n_p$. Assuming for simplicity that all plaquettes have the same size $n_p=n$ and that the wave function contains $M$ of them, the number of variational parameters $C_p(\bm \sigma_p)$ is $M \times 2^{n}$. If non overlapping (i.e., disentangled) plaquettes with $n > 1$ are considered, one recovers a cluster mean-field Ansatz in which inter-plaquette correlations and entanglement are fully neglected (see, e.g., Fig.~\ref{Fig:1}(a) for plaquettes of 2 adjacent sites). The crucial aspect of EPS is that plaquettes can be overlapped (i.e., entangled), introducing entanglement and correlations over distances larger than the plaquette size. 

Most of the literature on EPS \cite{Mezzacapoetal2009, MezzacapoC2010, Al-Assametal2011, MezzacapoB2012, Mezzacapo2012, Duricetal2014, Duricetal2016,Stojevicetal2016} has considered Ans\"atze based on { $M\leq N$} overlapping plaquettes of $n$ adjacent sites ($n$A-EPS), and improved them systematically by increasing $n$ (Fig.~\ref{Fig:1}(b,c)). While the $n$A-EPS Ansatz is asymptotically exact in the large $n$ limit, and any finite-$n$ A-EPS is a valid variational choice, convergence of the results in the plaquette size is often not achievable, as a result of the mentioned exponential cost of increasing $n$. This aspect becomes particularly serious in the vicinity of a quantum phase transition. For instance, as numerically found in Ref.~\cite{Stojevicetal2016}, an $n$A-EPS Ansatz with partially overlapping uniform plaquettes generally exhibits a finite correlation length (proportional to $n$), preventing the correct description of the critical regime at fixed $n$. In general, the extrapolation of the results to the large-$n$ limit is necessary to correctly determine the  ground state phase boundaries of a given model of interest \cite{Mezzacapo2012, MezzacapoB2012}.       

An extremely simple alternative to the above scheme consists in fixing the size of the plaquettes while playing with their geometry, which can be promoted to include (clusters of) sites  at arbitrary distances \cite{Changlanietal2009, Al-Assametal2011}. In this way correlations can be established directly among arbitrarily distant sites. This is the strategy underlying the long-range EPS Ansatz that we  consider here  in order to tackle systems that develop a diverging correlation length in their ground state. In particular, we shall focus on plaquettes with an even number of sites, $n$, and establish  their geometry as composed of \emph{two} clusters of $n/2$ adjacent sites at an arbitrary distance from each other: this ultimately defines the $n$LR-EPS wave function as illustrated in Fig~\ref{Fig:1}(d,e). 
The $n$LR-EPS clearly extends the $n$A-EPS, explicitly incorporating dominant $n$-site correlations at all length scales. Such an improvement occurs at a computational price now independent of $n$, and polynomial in the system size. 

 In general, the $n$LR-EPS Ansatz is systematically improvable upon increasing $n$, similarly to the A-EPS Ansatz. However, unlike in the A-EPS Ansatz, this improvement procedure is not  \emph{a priori} crucial, since the long-range physics of a given Hamiltonian can be well captured already with $n=2$ (see Secs.~\ref{s.XX} and \ref{s.J1J2}).  Another strategy for the improvement of the Ansatz that we shall pursue here is to combine both adjacent-site as well as long-range plaquettes, to give rise to $n$LR-$n'$A-EPS (where $n$ ($n'$) is the size of the long-range (adjacent-site) plaquettes, with $n'>n$)  exemplified in Fig.~\ref{Fig:1}(f). The wave-function coefficients of this Ansatz are defined via:
 \begin{equation}
 \psi_{n{\rm LR}-n'{\rm A-EPS}}(\bm \sigma) =  \psi_{n{\rm LR-EPS}}(\bm \sigma) ~\psi_{n'{\rm A-EPS}}(\bm \sigma)~.
\end{equation}
The latter strategy allows one in principle to  faithfully describe $n$ point correlations at all distances, also improving on local properties such as the energy of short-range interacting models.    
 
 In the particular case of the 2LR-EPS Ansatz and $S=1/2$ spins, for a $N-$site lattice { $M=N(N-1)/2$} plaquettes are formed by each pair of sites $i$ and $j$, i.e., $p=(ij)$ in Eq.~\eqref{e.EPS}, and the wave function coefficients can be explicitly rewritten as 
 \begin{eqnarray}
 \psi_{\rm 2LR-EPS}(\bm \sigma) & = & \prod_{i<j} C_{ij}(\sigma_i,\sigma_j), \nonumber \\
C_{ij}(\sigma_i,\sigma_j)  &=& \exp(a_{ij}+b^{(1)}_{ij} \sigma_i + b^{(2)}_{ij} \sigma_j + c_{ij} \sigma_i \sigma_j)~~
 \label{e.2LR-EPS}
 \end{eqnarray}
 establishing a link with what goes under the name of spin-Jastrow Ansatz in the previous literature \cite{Pang1991}.  Therefore, our approach generalizes systematically spin-Jastrow states, both in terms of the size of the plaquettes that are explicitly correlated in the form of the Ansatz; as well as via the use of complex plaquette coefficients, [see detail in Sec.~\ref{s.complex}].  It goes without saying that the EPS Ansatz applies to any lattice geometry and to any number of spatial dimensions without requiring particular modifications. The explicit correlations introduced within all pairs of clusters allows one to describe long-range entanglement in any such situation.  
  
\subsection{Variational energy minimization}  
  
Given the Hamiltonian ${\cal H}$, the variational optimum is searched via imaginary-time evolution projected onto the space of states compatible with the variational Ansatz via the time-dependent variational principle \cite{bookTDVP} - this approach is equivalent to the so-called stochastic reconfiguration scheme \cite{BeccaS2017, CarleoT2017}. The variational energy to be minimized reads
\begin{equation}
\langle {\cal H} \rangle = \langle E(\bm \sigma; \bm C) \rangle = \sum_{\bm \sigma} E(\bm \sigma;\bm C) P(\bm \sigma;\bm C)
\label{e.meane}
\end{equation}   
where 
\begin{equation}
P(\bm \sigma;\bm C) = \frac{|\psi(\bm \sigma;\bm C)|^2}{\sum_{\bm \sigma'} |\psi(\bm \sigma';\bm C)|^2}
\end{equation}
 and $E(\bm \sigma; \bm C) $ is the energy estimator
\begin{equation}
E(\bm \sigma;\bm C) = \sum_{\bm \sigma'}\langle \bm \sigma | {\cal H} | \bm \sigma' \rangle \frac{\psi(\bm \sigma';\bm C)}{\psi(\bm \sigma;\bm C)}  
\label{e.energy}
 \end{equation} 
 Introducing the logarithmic derivatives with respect to the $l$-th variational parameters of the wave function coefficients
 \begin{equation}
 L_l(\bm \sigma; \bm C)=\frac{1}{\psi (\bm \sigma; \bm C)}\frac{\partial\psi(\bm \sigma; \bm C)}{\partial C_l}~,
 \end{equation}
 the gradient
 \begin{equation}
 g_l=\langle L_l^* E \rangle-\langle L_l^* \rangle \langle E \rangle,
  \label{e.gradient}
 \end{equation} 
 as well as the covariance matrix
 \begin{equation}
 S_{l,m}=\langle L_l^* L_m \rangle-\langle L_l^* \rangle \langle L_m \rangle~,
 \label{e.Smatrix}
 \end{equation}
 the projected imaginary-time dynamics of the variational Ansatz is then described by the equation 
 \begin{equation}
 \dot{\bm C} = - \bm S^{-1} {\bm g}~.
 \label{e.TDVP}
 \end{equation} 
 The simple form of the EPS Ansatz in Eq. \eqref{e.EPS} allows for a straightforward calculation of the $L_l$'s, whose expression reduces, for $\bm C=\{C_p(\bm \sigma_p)\}$, 
 to 
 \begin{equation}
 L_{C_p(\bar{\bm \sigma}_p)}=\frac{\delta_{\bm \sigma_p,\bar{\bm \sigma}_p}}{C_p(\bm \sigma_p)}~.
 \end{equation}
 Statistical sums over the configurations $\bm \sigma$, contained in Eqs.~\eqref{e.meane}, \eqref{e.gradient} and \eqref{e.Smatrix}, are sampled via a Monte Carlo scheme which makes use of  Metropolis updates based on the exchange of spins with opposite $\sigma_z$. Hence, the total spin along $z$, set to $0$ in the initial state, is conserved along the simulation. Further details on optimization strategies  will be described in the next paragraph.

\subsection{Complex wave function coefficients}   
\label{s.complex} 
  
 In order to search for the ground state of frustrated magnetic models with a real-valued Hamiltonian, it is crucial to be able to account for weights $\psi(\bm \sigma)$  of the wave function with positive as well negative sign \cite{BeccaS2017}. The function ${\rm sgn}[\psi(\bm \sigma)]$ defines the so-called \emph{sign structure} of the state, and it is in general unknown.  In fact, one can argue that knowing it \emph{a priori} would result in being able to solve the ground state problem by using QMC techniques based on ground-state projection (such as Green's function Monte Carlo \cite{BeccaS2017}) with fixed-node constraints. Indeed, the sign structure of the ground state is known to be trivial (all positive weights) when the off-diagonal elements of the Hamiltonian matrix $\langle \bm \sigma' | {\cal H} | \bm \sigma \rangle$ are semi-negative definite (Perron-Frobenius theorem), or can be made so by an unitary transformation. This is the case for instance of antiferromagnetic Hamiltonians defined on a bipartite lattice, namely containing only antiferromagnetic interactions between sites belonging to different sublattices ($A$ and $B$): in that case the unitary transformation (amounting to a $\pi$-angle spin rotation on all $A$ spins) leads to a sign structure 
determined by the  Marshall sign rule \cite{Auerbachbook}: ${\rm sgn}[\psi(\bm \sigma)] = (-1)^{N_{A,\uparrow}(\bm \sigma)}$ (where $N_{A,\uparrow}(\bm \sigma)$ is, for the generic configuration $\bm \sigma$, the number of $\uparrow$ spins on the $A$ sublattice). 
 
 In order to tackle generic systems, in which the ground-state signs are completely unknown, an ideal variational Ansatz should therefore be able to reproduce a wide variety of possible sign structures, and to do so in a continuous manner, so that the sign structure can be a subject of variational optimization.  In the EPS Ansatz the sign of  $\psi(\bm \sigma)$ is simply given by the product of the signs of the plaquette coefficients $C_p(\bm \sigma_p)$, introducing a fundamental bias on the possible sign structures that the Ansatz can realize. Nonetheless it is a simple exercise to prove that the Marshall sign structure can indeed be reproduced by the EPS Ansatz, as discussed in Appendix \ref{a.Marshall}. 
 From the point of view of the numerical optimization,  an EPS Ansatz with  \emph{real}   plaquette parameters (obeying the dynamics governed by the equation Eq.~\eqref{e.TDVP}) should in principle lead to changing the signs of the wave function weights, so as to explore non-trivial sign structures in a purely variational way. However, in practice,  a real coefficient $C_p(\bm \sigma_p)$ approaching zero (necessary for a sign change) entails that the weight $\psi(\bm \sigma)$ of  all the configurations $\bm \sigma$ compatible with $\bm \sigma_p$ on the $p$-th plaquette also approaches zero, due to the multiplicative structure of the EPS Ansatz. This in turn makes the appearance of the $\bm\sigma_p$ configuration in the update very rare, so that an enormous statistics has to be accumulated in order to properly sample the corresponding gradient (Eq.~\eqref{e.gradient}) and covariance matrix (Eq.~\eqref{e.Smatrix}). A very simple strategy to circumvent this issue is to extend the weights to the complex plane by considering complex-valued $C_p(\bm \sigma_p)$ so that sign changes (or $\pi$ phase shifts) can be made without ever crossing the origin: this allows then for a full variational optimization of the \emph{phase} structure of the Ansatz, which is the strategy that we pursue in this study. The use of complex  $C_p(\bm \sigma_p)$ should be seen as an extension of the variational space in order to be able to recover the correct variational optimum which, for a real-valued Hamiltonian matrix, should also be real up to a global phase.  
 Our variational optimization procedure indeed leads to states consistent with this scenario.
 
 From a technical point of view, it is convenient to parametrize the plaquette coefficients in their polar decomposition $C_p(\bm \sigma_p) = A_p(\bm \sigma_p)
 e^{i\theta_p(\bm \sigma_p)}$ This allows, for example, to optimize the amplitude $A$ and phase $\theta$ as independent variables. Also, one can optimize the phases at first while keeping the amplitudes fixed and equal to one: in doing so one produces a  Monte Carlo dynamics for the sampling of the statistical sums of Eqs.~\eqref{e.meane}, \eqref{e.gradient} and \eqref{e.Smatrix}, where all configurations are equally probable. After a possible pre-optimization of the phases (achieved when the variational energy ceases to decrease), the amplitudes are left free to vary (i.e., to depart from their initially { unit} value) and optimized alongside with the phases. In this way any nodal and sign structure compatible with the Ansatz may emerge from the optimization in an unbiased fashion. Moreover, the introduction of spatial symmetries in the plaquette parameters can be done separately for the amplitude and phase variables, namely the functions $A_p$ and $\theta_p$ can be made to depend on the plaquette $p$ index in a different way. Explicit examples will be provided in Sec.~\ref{s.J1J2}.
 
 \subsection{Correlation functions and entanglement entropies}
 
  The main focus of our present work is on the ability of the nLR-EPS Ansatz to correctly capture the correlation and entanglement properties of quantum spin states with small (or even minimal, \emph{i.e.}, $n=2$) plaquette size. All of the findings presented in this study concern one-dimensional $S=1/2$ quantum spin models, described by spin operators $S_i^{\alpha}$, where $\alpha = x, y, z$ and the index $i$ runs on the lattice sites of a linear chain.  
 In the following we shall present results for the spin-spin correlation function   
 \begin{equation}
 C^{\alpha\alpha}(r) = \frac{1}{N} \sum_i \langle S^\alpha_i S^\alpha_{i+r}\rangle
 \end{equation}
 and the related structure factor 
 \begin{equation}
 {\cal S}^{\alpha\alpha}(k) = \frac{1}{N} \sum_r C^{\alpha\alpha}(r) e^{ikr};
 \end{equation}
 as well as for the dimer order paremeter defined as in Ref.~\onlinecite{Sandvik2010} as
 \begin{equation}
 \mathcal{D}_N=[D(N/2)-D(N/2-1)
]/2
\label{e.DOP}
 \end{equation} 
 where 
  \begin{equation}
 D(r) = \frac{1}{N} \sum_i \left [ \langle (\bm S_i \cdot \bm S_{i+1})(\bm S_{i+r} \cdot \bm S_{i+r+1})\rangle - \langle \bm S_i \cdot \bm S_{i+1}\rangle^2\right ]~.
 \end{equation}
is the dimer-dimer correlation function.
 Moreover we will concentrate on the 2-R\'enyi entanglement entropy $R_2(A)=-\log({\rm Tr} \rho^2_A)$ where $\rho_A = {\rm Tr}_B |\Psi\rangle \langle \Psi|$ is the reduced density matrix describing the subsystem $A$ after having traced out the degrees of freedom of its complement $B$. The purity ${\rm Tr} \rho^2_A$ can be conveniently calculated as the expectation value of the SWAP operator of the configuration of the $A$ subsystem between two replicas of the whole system \cite{Hastingsetal2010}. Denoting with  $\bm \sigma_A$ and $\bm \sigma_B$ the configurations of the subsystems $A$ and $B$ in a state $\bm \sigma = (\bm \sigma_A, \bm \sigma_B)$ of the computational basis, one has 
 \begin{equation}
 {\rm Tr} \rho^2_A = \langle {\rm SWAP}_A\rangle_2 = \left \langle \frac{\psi(\bm \sigma'_A,\bm\sigma_B)\psi(\bm \sigma_A,\bm \sigma'_B)}{\psi(\bm \sigma_A,\bm\sigma_B)\psi(\bm \sigma'_A,\bm \sigma'_B)} \right \rangle_2
 \end{equation}
 where the two-replica statistical average $\langle ... \rangle_2$  is defined as 
 \begin{equation}
\langle ... \rangle_2  = \sum_{\bm \sigma_A,\bm \sigma_B} \sum_{\bm \sigma'_A,\bm \sigma'_B} |\psi(\bm \sigma_A,\bm \sigma_B)|^2 |\psi(\bm \sigma'_A,\bm \sigma'_B)|^2 (...)~.
\end{equation}

  \subsection{Alternative variational Ans\"atze for quantum spin models and comparison with LR-EPS}
  \label{s.review}

   In this section we briefly review some of the most popular variational states for lattice spin models, and contrast their properties with those of our LR-EPS Ansatz. 
   
  The most successful example of a variational Ansatz is represented by matrix-product states (MPS), which represent the variational Ansatz optimized by the density-matrix renormalization group (DMRG) algorithm as well as other related techniques \cite{Schollwoeck2005}. The MPS capture with impressive precision the physics of one-dimensional quantum systems with entanglement entropies obeying an area-law scaling with subsystem size, including possible logarithmic corrections. Indeed the maximum sub-system entanglement entropy that the Ansatz can allow for is given by $\log D$ where $D$ is the linear dimension of the matrices composing the Ansatz and allows for its systematic improvement. In this framework, only logarithmic scalings of the entanglement entropy with subsystem size are tolerable in order to achieve a polynomial scaling of the number of variational parameters with system size.  
  
    The direct application of MPS to models in higher spatial dimension is in principle problematic. In fact, in this case, the $D$ parameter scales exponentially with the subsystem size even in the case of area-law states \cite{StoudenmireW2012}. This issue can be circumvented by generalizing MPS to tensor-network states (TNS) \cite{Orus2014}, which are suited to study the physics of area-law states in two and higher dimensions. Like MPS, TNS can be systematically improved by enlarging the number of parameters with a concomitant polynomial scaling in the computational cost; nonetheless TNS generally exhibit a finite correlation length \cite{Corbozetal2018}; in several formulations (such as the celebrated projected entangled-pair states \cite{Verstraeteetal2008}) they do not allow for an efficient exact calculation of the wave function coefficients \cite{Glasseretal2018} nor for an efficient representation of quantum states with faster than area-law entanglement scaling.   
  Many Ans\"atze offer valuable alternatives to MPS/TNS with rather complementary properties. A famous example is given by resonating valence-bond (RVB) states \cite{Liangetal1988}, efficiently parametrized as projected Bardeen-Cooper-Schrieffer (pBCS) states \cite{Anderson1987}. Their sign structure descends from that of a fermionic determinant which provides the ground state of a BCS-like Hamiltonian, and as such it can be highly non-trivial. {As a consequence the pBCS states have been successfully applied to several frustrated  models of quantum magnetism \cite{BeccaS2017}; nonetheless they cannot be systematically improved in their sign structure -- although one may argue that they could be combined \emph{e.g.} with complex-valued Jastrow factors altering both the amplitude and sign structure of the coefficients.} More recently the ability of neural networks to reproduce a function of many variables has been exploited to parametrize in this form the weights $\psi(\bm \sigma;\bm C)$, defining the Ansatz called neural-network quantum states (NNQS) \cite{CarleoT2017}. The latter proves to be very effective both for unfrustrated \cite{CarleoT2017} as well as frustrated  models of magnetism, with a fixed sign structure \cite{Chooetal2019} or by using complex coefficients \cite{Glasseretal2018}. Ref.~\cite{Caietal2018} recently explored the possibility of parametrizing the signs by using a dedicated neural network, but without complexification of the coefficients.  
NNQS as formulated in Ref.~\cite{CarleoT2017} can be systematically improved by increasing the depth of the neural network, although networks beyond single-layer ones do not allow for an efficiently calculable form of the coefficients (see Ref.~\cite{Chooetal2019} for a multi-layer convolutional network representation).
  LR-EPS with complex coefficients appear as a valuable alternative to all the above variational schemes, because of the relative simplicity of their structure and the flexibility of their formulation, with correlations and long-range entanglement explicitly built in. EPS come with a large variety of improvement schemes (e.g., extension of the plaquette sizes, combination of long-range and adjacent-site plaquettes) as well as with the possibility to achieve a full variational optimization of the sign structure (within the structures compatible with the Ansatz) by generalizing it to the phase structure of a complex wave function, as described above.

\section{Validation of the Ansatz: XX chain}
\label{s.XX}

A first validation of our approach comes from the case of the $S=1/2$ XX chain, with Hamiltonian
 \begin{equation}
 {\cal H} = - J  \sum_i \left ( S_i^x S_{i+1}^x + S_i^y S_{i+1}^y \right ) 
 \end{equation}

\begin{figure}[ht!]
\centerline{\includegraphics[width = 1.0\columnwidth]{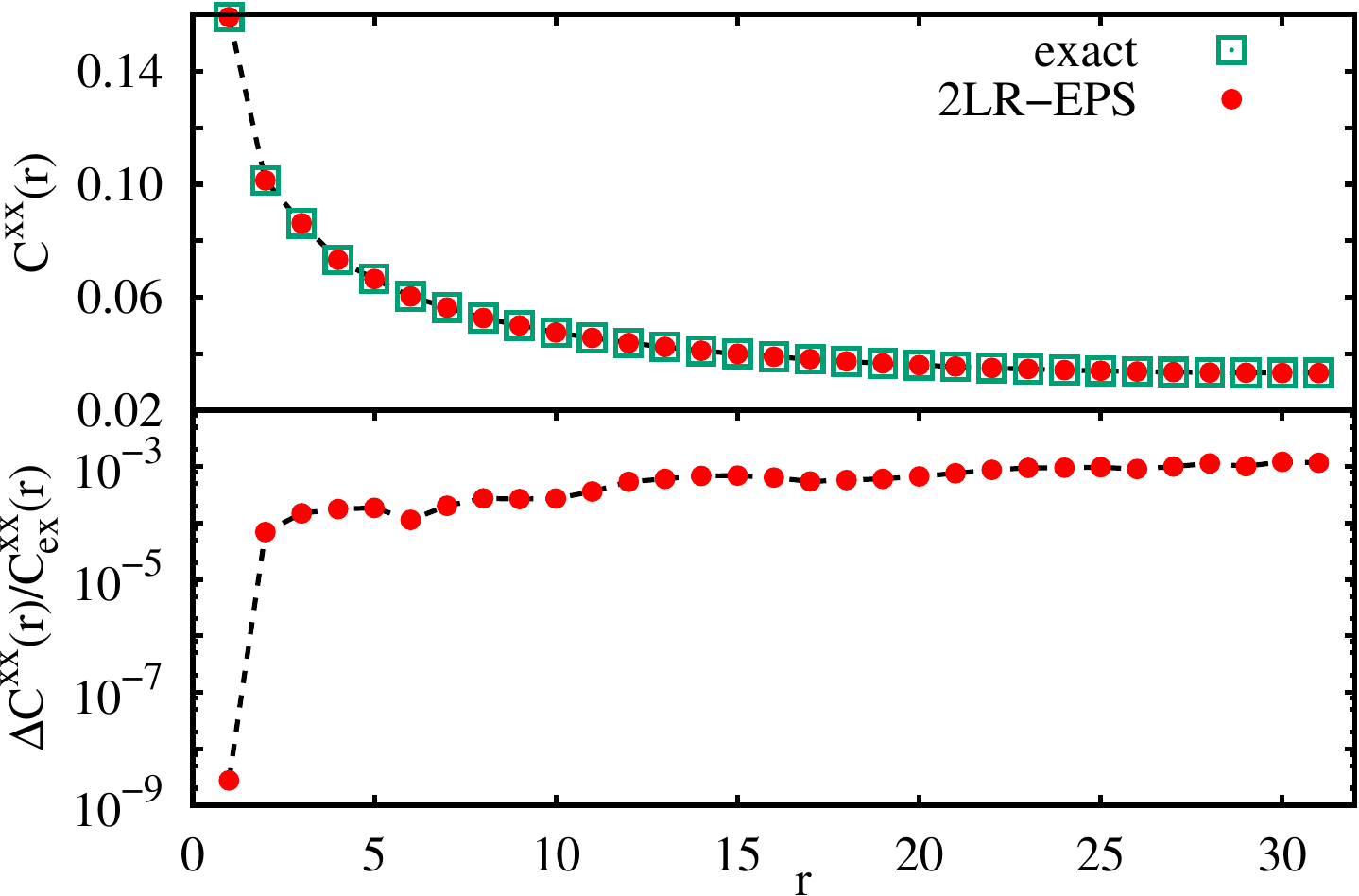}}
\caption{Ground state spin-spin correlation function of an XX chain {with $N=62$ sites} (upper panel). The error of the 2LR-EPS estimates relative to the exact ones is shown in the lower panel. Lines are guides to the eye} 
\label{Fig:2}
\end{figure}

\begin{figure}[t]
\centerline{\includegraphics[width = 1.0\columnwidth]{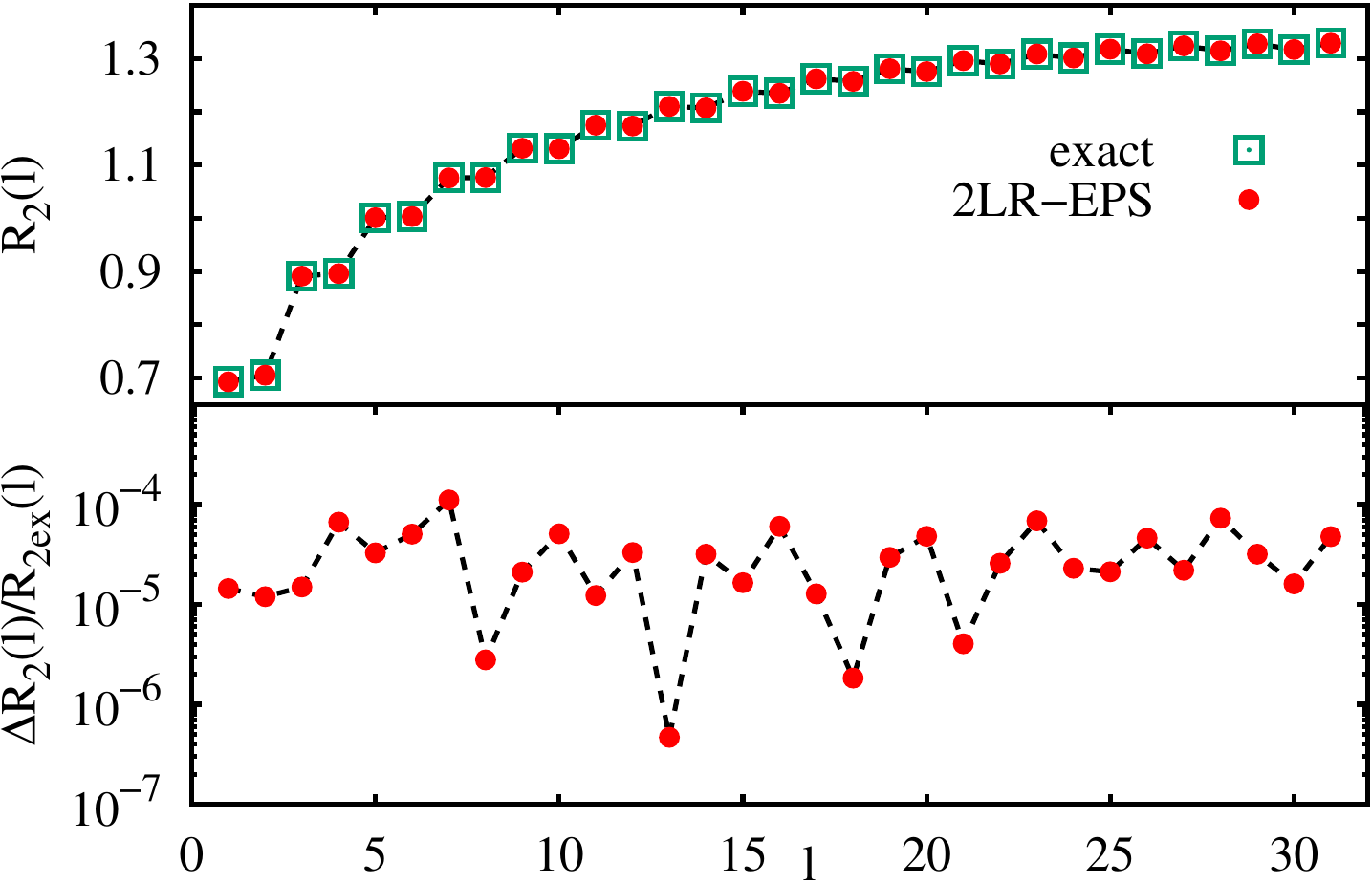}}
\caption{Ground state  2-R\'enyi entanglement entropy as a function of the subsystem size $l$ for a XX chain {with $N=62$ sites} (upper panel). The error of the 2LR-EPS estimates relative to the exact ones is shown in the lower panel. Lines are guides to the eye.} 
\label{Fig:3}
\end{figure}

  which is exactly solvable by mapping it to free fermions via the Jordan-Wigner transformation \cite{Liebetal1961}. Here and throughout the rest of the paper we consider chains with periodic boundary conditions. This Hamiltonian is not frustrated and satisfies the Perron-Frobenius theorem, so that its ground state wave function has all positive-defined coefficients. One can therefore search for its ground state in the form of a LR-EPS with real coefficients. In the context of LR-EPS, the case of the XX chain is in fact extremely special, because  the exact coefficients of the fermionic ground state, given by the Slater determinant, can be recast, in the form of a Vandermonde determinant \cite{Sutherland-book}, as
  \begin{equation}
  \psi_{XX}({\bm \sigma}) = {\cal N} \prod_{i<j} \sin\left[ \frac{\pi}{N} (j-i) \right]^{(2\sigma_i+1)(2\sigma_j+1)/4}
  \end{equation}  
  where ${\cal N}$ is a normalization factor. It is immediate to recognize that the above espression is fully compatible with that of the 2LR-EPS Ansatz, Eq.~\eqref{e.2LR-EPS}. In fact the 2LR-EPS form encompasses an entire family of many-body wavefunctions, such as the exact ground state of one-dimensional bosons or fermions with inverse-square interaction potential \cite{Sutherland1971}, the exact ground state of the Haldane-Shastry model \cite{Haldane1988,Shastry1988}, or the Laughlin wavefunction \cite{Changlanietal2009}, to cite a few relevant examples.  
  
  In the specific case of the XX chain, the ground-state physics features a critical Luttinger-liquid phase with algebraically decaying correlations $C^{xx(yy)} \sim r^{-1/(2K)}$ with Luttinger liquid exponent $K=1$ \cite{Giamarchi-book}, and a logarithmic scaling of the entanglement entropy of a subsystem of linear size $l$:
 \begin{equation}
  R_{2}(l) =  \frac{c}{4} \log(l) + c_1 + ...
  \label{e.R2}
 \end{equation}
 with $c_1$ a non universal constant and $c=1$ giving the central charge of the related conformal field theory \cite{CalabreseC2004} (the missing terms are subdominant corrections). These traits are in fact universal to many one-dimensional systems admitting a description as a Luttinger liquid \cite{Giamarchi-book}, whose scale-free, critical nature is nicely captured by the spatial structure of correlations inscribed in the LR-EPS Ansatz. Indeed Jastrow wave functions (to which a real-valued 2LR-EPS can be specialized) have already been assessed in the past as very efficient descriptions of Luttinger-liquid phases of both interacting bosons and fermions \cite{Capelloetal2005,Capelloetal2008}. 
 
 Here we use the simplest form of the Ansatz, namely a 2LR-EPS where we enforce translational symmetry, so that only ${\cal O}(N)$ variational parameters are required. A very high accuracy can be obtained for both the off-diagonal spin-spin correlation function $C^{xx}(r)$ [Fig.~\ref{Fig:2}] and the 2-R\'enyi entropy [Fig.~\ref{Fig:3}]. In particular it is remarkable to see that, for the energy, the error relative to the exact results is less than $10^{-8}$; in fact slightly lower than the statistical error that we have on the energy estimator itself. This means that, for all purposes, our optimization algorithm finds the absolute minimum of the variational energy, and that our accuracy on the reconstruction of the ground state is only limited by the statistical uncertainty.  The relative accuracy of the LR-EPS predictions remains rather good ($\lesssim 10^{-3}$ for correlations, $\lesssim 10^{-4}$ for the entanglement entropy) when looking at long-distances, without any significant degradation of our estimates being observed for increasing $r$, or $l$.

\section{$\bm{J_1-J_2}$ chain}
\label{s.J1J2}

\subsection{Model and phase diagram}
\label{s.J1J2generalities}
 
A more stringent test of our Ansatz is provided by the frustrated antiferromagnetic $J_1-J_2$ chain whose Hamiltonian is
\begin{equation}
{\cal H} = J_1 \sum_i {\bm S}_i \cdot {\bm S}_{i+1} + J_2 \sum_i  {\bm S}_i \cdot {\bm S}_{i+2},
\label{h:2}
\end{equation} 
This model features a very rich phase diagram upon changing the degree of frustration $\alpha = J_2/J_1$. At low frustration, its  ground state realizes a gapless phase (described as a Luttinger liquid) continuously connected to its unfrustrated limit $J_2=0$. When $\alpha = \alpha_c =  0.241167 ....$ \cite{Eggert1996} the system undergoes a quantum phase transition to a gapped, valence-bond crystal (VBC) with spontaneous dimerization. Beyond the exactly solvable Majumdar-Ghosh  (MG) point $\alpha = 1/2$ \cite{MajumdarG1969}, correlations start developing an incommensurate structure \cite{Bursilletal1995, WhiteA1996, Gehringetal1997, DeschnerS2013} with a pitch vector that evolves continuously towards $\pi/2$ as $\alpha$ increases. After its seminal numerical investigations via DMRG \cite{Bursilletal1995, WhiteA1996, Gehringetal1997}, Eq.~\eqref{h:2} has been successfully investigated in the more recent past with RVB states \cite{Ferrarietal2018}, as well as with NNQS \cite{Caietal2018}. 

In order to capture the rich phenomenology correctly with a variational approach, it is crucial to use a wave function with a flexible, continuously adjustable sign structure. 
This is clearly evidenced by the fact that correlation functions can exhibit different sign patterns, going from the staggered one (related to a pitch vector $\phi = \pi$) for small $\alpha$ to incommensurate sign patterns for $\alpha > 0.5$. Indeed the sign pattern of off-diagonal correlations $C^{xx(yy)}(r)$ is related to the mutual signs of coefficients corresponding to configurations connected by two spin flips at distance $r$. Indeed,
\begin{equation}
\langle S_i^+ S_{i+r}^- + S_{i}^- S_{i+r}^+ \rangle = \sum_{\bm \sigma} \psi^*(\bm \sigma) \psi(\bm \sigma_{i,i+r})
\end{equation}
where $\bm \sigma_{i,i+r}$ corresponds to the $\bm \sigma$ configuration with flipped $i$-th and $(i+r)$-th spins.  Therefore, off-diagonal correlations with a non-trivial sign pattern are consequence of the sign structure of the wave function coefficients. A closer look at the ground-state sign structure of model \eqref{h:2}  has been offered in Ref.~\cite{ZengP1995}, pointing out that a Marshall sign for a two-sublattice structure $ABAB...$ appears at weak frustration $\alpha \ll 1$, while a Marshall-like sign with sublattice structure $ABBA...$ emerges in the opposite limit $\alpha \gg 1$.  For intermediate values $\alpha \sim {\cal O}(1)$ a definite sign structure could not be identified, and most likely it evolves with $\alpha$ along with the incommensurate correlations.

It is instructive to remind the reader that the mean-field (MF) solution of the problem gives a wavefunction 
\begin{equation}
|\Psi_{\rm MF}\rangle = \otimes_j \left ( |\uparrow_j\rangle + e^{iQ r_j} |\downarrow_j\rangle \right )
\end{equation}
where $Q = \pi$ for $\alpha < 1/4$ and $Q = \cos^{-1}[-1/(4\alpha)]$ for $\alpha > 1/4$. Therefore dressing the MF wavefunction with a real correlation term {(\emph{e.g.} of the Jastrow type)} would already produce non-trivial incommensurate correlations. Nonetheless the exact solution of the ground-state problem of a time-reversal invariant Hamiltonian (such as the one under investigation) is given by a state $|\Psi_0\rangle$  that can only possess a time-reversal invariant total momentum -- namely, under a translation $T_\delta$ of a distance $\delta$, $T_\delta |\Psi_0\rangle = e^{iP\delta} |\Psi_0\rangle$  with $P = 0$ or $\pi$.  {On the other hand, projecting the above mean-field state onto the sector at zero total magnetization ${\cal P}_{S^z=0}|\Psi_{\rm MF}\rangle$ produces a state which has a momentum $P = NQ/2$ (because upon translation of $\delta$ sites the $N/2$ $\downarrow$ spins produce a phase factor $\exp(i NQ\delta/2)$). This situation, however, results in a ground-state wave function leading  to both incommensurate correlations and an unphysical breaking of the time-reversal symmetry (i.e., the ground state remains complex-valued). Conversely, since our optimized  Ansatz is essentially real (see discussion in Sec.~\ref{a},~\ref{aa}, and Appendix B) the emergence of incommensurate correlations discussed in this work cannot be due to the above scenario.

\subsection{Validation of the Ansatz: comparison with exact results}
\label{a}
  In light of the above observations, the 1$d$ $J_1-J_2$ model offers a rather challenging testbed for our variational Ansatz. In the following we shall check the accuracy of our results against exact ones by comparing ground-state energies and correlation functions. First of all, in the case of the unfrustrated Heisenberg chain ($J_2=0$) the optimization of complex-valued LR-EPS reproduces faithfully the physics of the system: when, for instance, the calculation starts from a state with random coefficients of unit norm (i.e., $C_p(\bm\sigma_p) = A_p(\bm\sigma_p)e^{i\theta_p(\bm\sigma_p)}$ with $\theta_p(\bm\sigma_p)$, and $A_p(\bm\sigma_p)$ initially chosen as a random phase between $0$ and $2\pi$, and $1$, respectively), the optimization algorithm is capable of finding efficiently the Marshall sign structure, as witnessed by the correct sign pattern reconstructed for the correlation function  $C(r)=\sum_{\alpha}C^{\alpha \alpha}(r)$ in Fig.~\ref{Fig:4}. This shows that the variational search of the sign structure, when compatible with the LR-EPS multiplicative form, can indeed be efficiently performed. 
Moreover, our estimates are in very good agreement with numerically exact QMC results based on the stochastic series expansion \cite{Sandvik2010}. We note that,  by means of a $n$A-EPS wave function explicitly including the Marshall signs, one may achieve for the present model and system size an accuracy on the ground state energy similar to that obtained with the 2LR-EPS, when $n$ is as large as $12-14$ sites (i.e., via a considerably larger number of variational parameters than in the 2LR-EPS case).

  \begin{figure}[t]
\centerline{\includegraphics[width = 1.0\columnwidth]{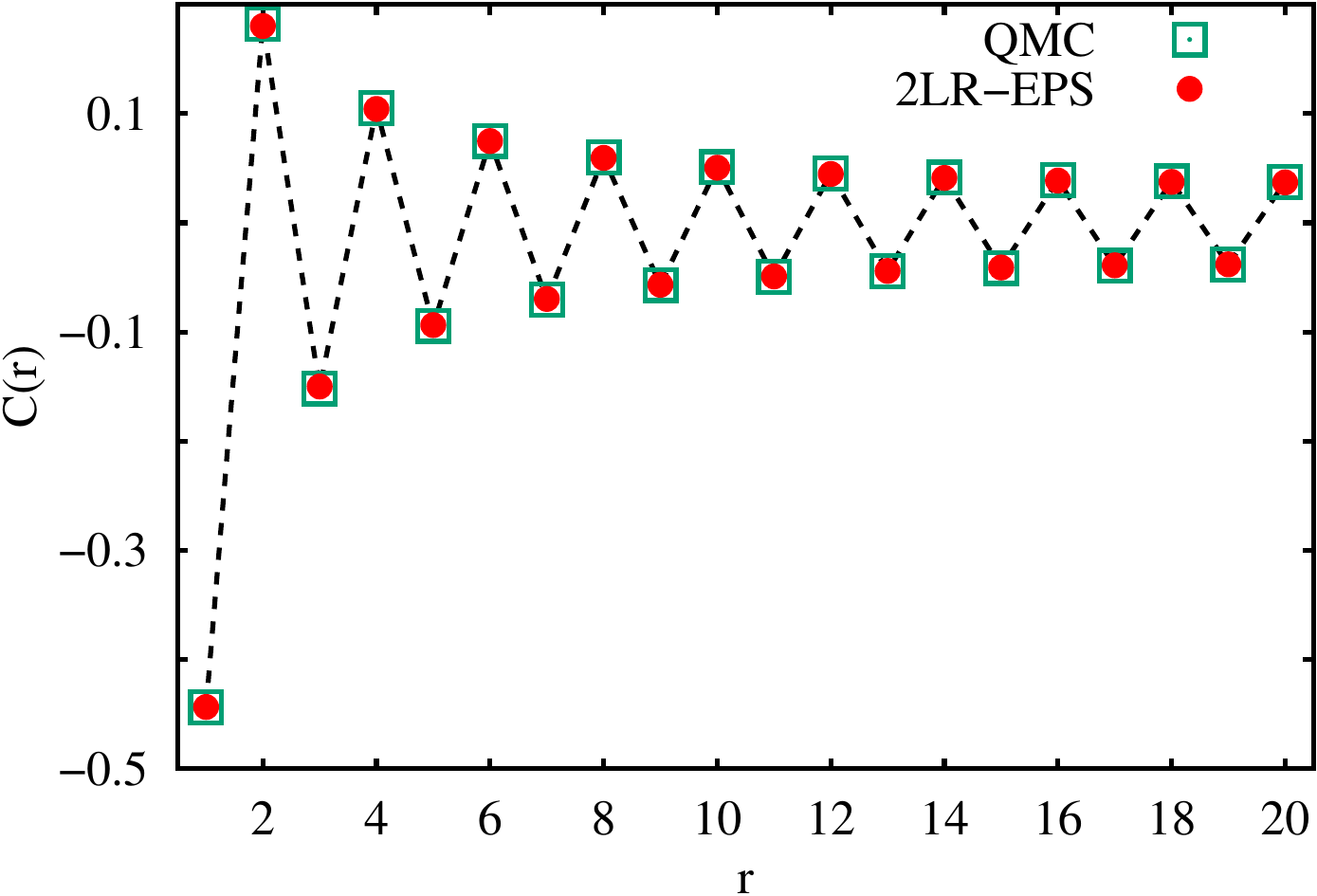}}
\caption{Ground state spin-spin correlation function for a  chain of $N=40$ sites governed by the $J_1-J_2$ Hamiltonian in the unfrustrated case (i.e., $J_2=0$). The dashed line is a guide to the eye.} 
\label{Fig:4}
\end{figure}

 \begin{figure}[t]
\centerline{\includegraphics[width = 1.0\columnwidth]{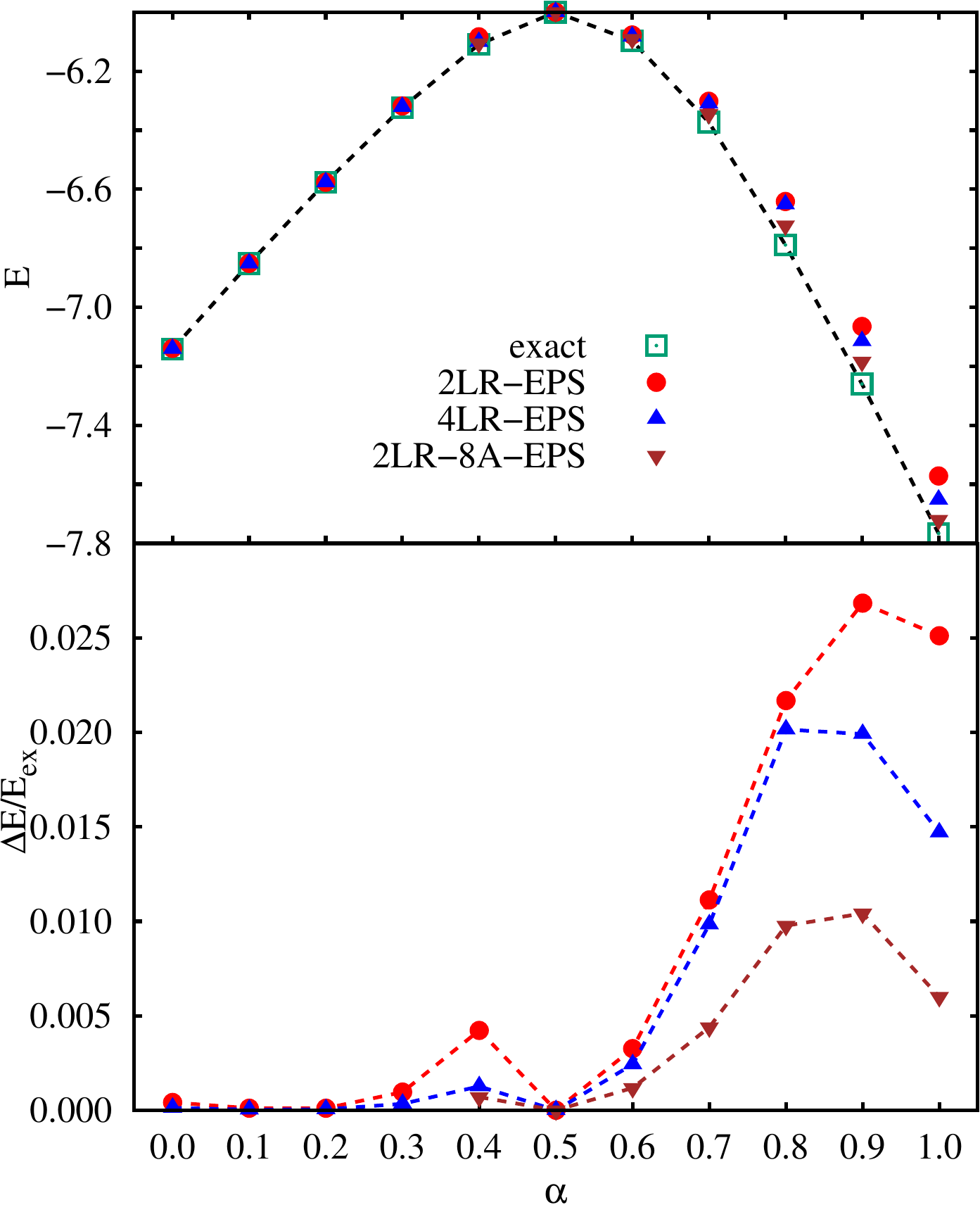}}
\caption{Ground-state energy estimates obtained with various EPS Ans\"atze (upper panel), and their error relative to the exact result (lower panel) as a function of $\alpha$. The system size is $N=16$. Lines are guides to the eye.} 
\label{Fig:5}
\end{figure}
\begin{figure}[t]
\centerline{\includegraphics[width = 1.0\columnwidth]{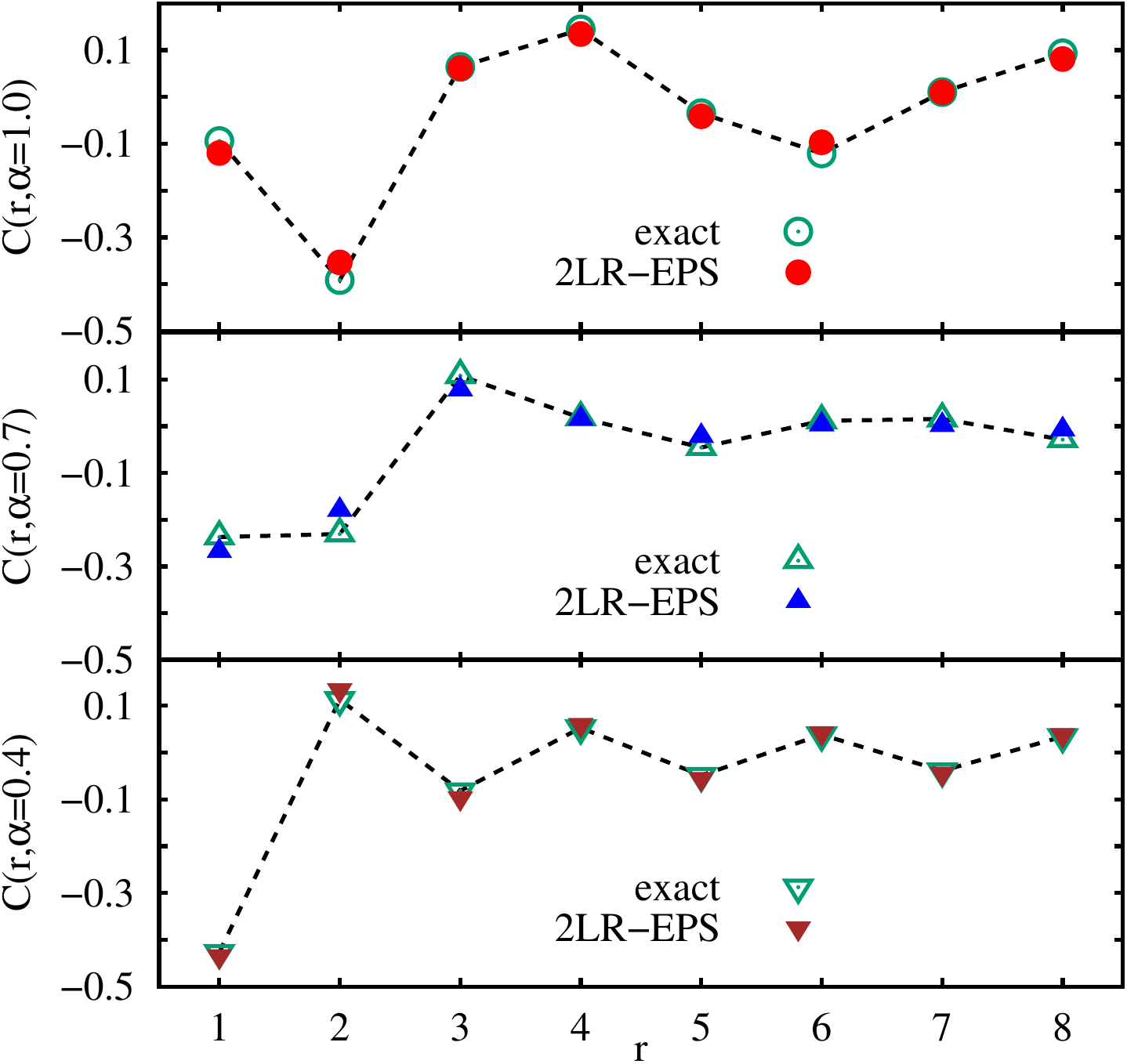}}
\caption{Ground-state spin-spin correlation function for a  chain of $N=16$ sites governed by the $J_1-J_2$ Hamiltonian, and for various values of the frustration parameter $\alpha$. Lines are guides to the eye.} 
\label{Fig:6}
\end{figure}  

 Results at finite frustration compared with exact diagonalization ones are reported in Fig.~\ref{Fig:5} and \ref{Fig:6}.  
In paricular, Fig.~\ref{Fig:5}~(upper panel) shows the $\alpha$ dependence of the variational energy of different LR-EPS states for a $N=16$ chain with periodic boundary conditions. The simplest Ansatz that we test is the 2LR-EPS one, where we do not impose any symmetry, namely we optimize $2N(N-1)$ independent complex coefficients. We observe that the 2LR-EPS Ansatz remains very accurate (with relative errors in the $0.1\%$ range)  up to  $\alpha \simeq 0.5$.  For larger values of $\alpha$ the precision degrades, with relative errors rising to the $1\%$ level) -- yet the Ansatz can be systematically improved by moving to the 4LR-EPS or to the mixed long-range/adjacent-site 2LR-8A-EPS Ansatz (i.e., by increasing, if no symmetries are considered, the number of variational parameter by a factor of about $3.5$ or $9.5$, respectively). Nonetheless the minimal 2LR-EPS wave function is already capable of reproducing the non-trivial and $\alpha$-dependent sign pattern of the correlations as shown in Fig.~\ref{Fig:6}. In our view, this offers valuable evidence of the fact that our simplest wavefunction can describe the different physical regimes of the model in question, without the need to further enrich its parametrization. Therefore we shall focus on this wavefunction for the study of larger system sizes, further reducing its variational parameters by imposing symmetries (see below).
 
 A further significant test of the quality of our results is offered by the real vs. complex nature of the wavefunction coefficients. Randomly choosing pairs of basis states $|\bm \sigma\rangle$  and $|\bm \sigma'\rangle$ we find that the ratio of their coefficients $\psi_{\rm 2LR-EPS}(\bm \sigma)/\psi_{\rm 2LR-EPS}(\bm \sigma')$ is a real number (whenever both coefficients are larger than machine precision in modulus). This is a direct proof that the variational optimization of the complex Ansatz is able to eventually align the phases of all the wavefunction coefficients (modulo $\pi$), returning a real valued wavefunction up to an irrelevant global phase. This offers additional evidence that the strategy of producing non-trivial sign structure by optimizing a complex wavefuction is successful.
       
\subsection{Correlations and entanglement on larger lattices} 

Having validated our Ansatz for the frustrated $J_1-J_2$ chain for a small lattice size, we move on to studying the physics of the same chain for larger lattices (up to $N=80$). Our goal is to show that the LR-EPS Ansatz is a very good tool to study fundamental features of the correlations and the entanglement of the ground state of the system, and their evolution across the phase diagram of the model. Throughout this section we specialize our attention to a 2LR-EPS Ansatz with coefficients $C_{ij}(\sigma_i,\sigma_j) = A_{ij}(\sigma_i,\sigma_j) e^{i\theta_{ij}(\sigma_i,\sigma_j)}$ where, without loss of generality, we consider $i < j$. We parametrize the spatial dependence of both amplitudes $A_{ij}$ and phases $\theta_{ij}$ in terms of the coordinate $i$ of the first site and of the distance between the sites $d= j-i$. In order to reduce the number of variational parameters, we chose the dependence on $(i, d)$ to be periodic of period $(p_i, p_d)$. All the results for lattices of size $N>16$ and in particular those in the following sections are obtained via wave functions with periods $(p_i, p_d)=(2,N)$ for the amplitudes (allowing to reproduce correlations at all distances, and a possible spontaneous dimerization of the lattice) and $(N,2)$ for the phases (allowing to describe relevant sign patterns with a number of phase parameters linear in $N$, as discussed in Appendix \ref{a.Marshall}). 
Further insight into the accuracy of the 2LR-EPS Ansatz on large lattices can be obtained by comparing our estimated ground state energies for the $J_1-J_2$ model with DMRG results. As an example, here we focus on a chain of $80$ sites and $\alpha=0.7$, i.e., in the parameter region where the Ansatz is quantitatively less accurate (see Fig.~\ref{Fig:5}). In the mentioned case we find that the 2LR-EPS Ansatz yields a ground state energy characterized by an error relative to the DMRG estimate \cite{Vodola} of approximately 0.76\%. When the 2LR-EPS Ansatz is improved by means of the 10A-EPS one resulting in the 2LR-10A-EPS Ansatz such a relative error decreases to approximately 0.37\%.

  \subsubsection{Spin-spin correlations in the gapless phase}
  Throughout the gapless phase $\alpha \leq \alpha_c$, $2$-point correlations exhibit a power-law decay as $d(r|N)^{-1}$ [with the chord length $d(r|N)=(N/\pi)\sin(\pi r/N)$]. The latter is the only one compatible with SU(2) symmetry for a gapless Luttinger liquid. Indeed, when mapping the spin model to a hardcore boson chain, the predictions of Luttinger liquid theory for the decay of the off-diagonal correlation function $\langle S^+_i S^-_{i+r}  \rangle=  \langle b_i^\dagger b_{i+r} \rangle \sim d(r|N)^{-1/(2K)}$,  and of the diagonal correlation function $\langle S^z_i S^z_{i+r}  \rangle=  \langle (n_i-1/2) (n_{i+r}-1/2) \rangle \sim d(r|N)^{-2K}$  \cite{Cazalilla2004} must coincide in the presence of SU(2) symmetry, hence the value $K=1/2$ for the Luttinger exponent. A more detailed analysis \cite{Eggert1996} points out the existence of a multiplicative logarithmic correction to the power-law decay, in the form $\sqrt{\log(r/r_0) \lambda_0}$, where $r_0$ and $\lambda_0$ are coefficients continuously depending on $\alpha$. Both coefficients vanish as one approaches the critical point with the constraint $-\lambda_0 \ln r_0 \to 1$, so that the logarithmic correction disappears in the same limit.
   \begin{figure}[b]
\centerline{\includegraphics[width = 1.0\columnwidth]{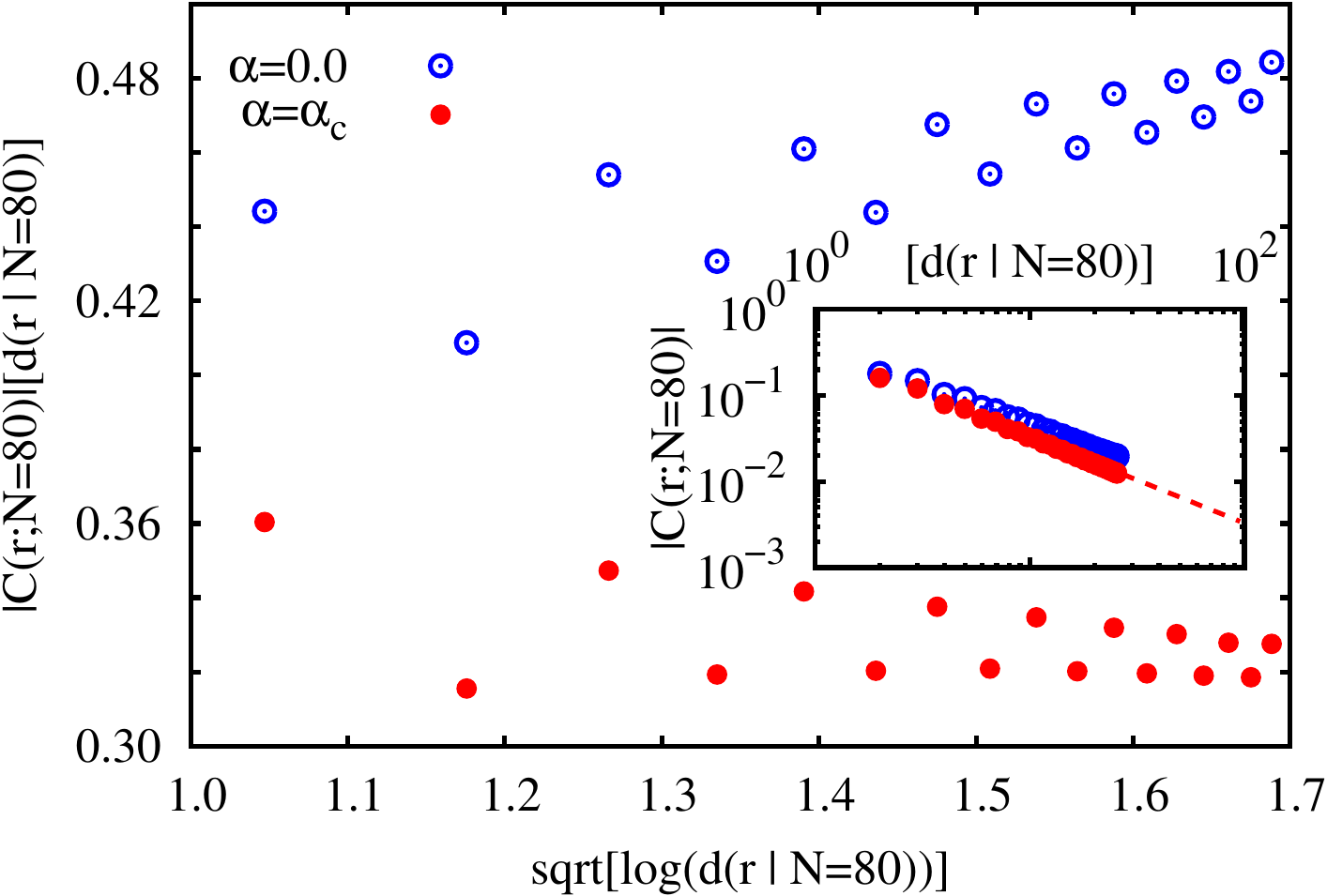}}
\caption{Absolute value of the ground-state spin-spin correlation function of the $J_1-J_2$ chain with $N=80$ sites $|C(r,N=80)|$ multiplied by the chord length $d(r|N=80)$, and plotted  as a function of the square root of the logarithm of the latter. Inset: $|C(r,N=80)|$ versus  $d(r|N=80)$; the  line is a fit to the corresponding numerical data (see text). Estimates are obtained with the 2LR-EPS Ansatz.} 
\label{Fig:7}
\end{figure}  
 Fig.~\ref{Fig:7} shows the absolute value of the spin-spin correlation multiplied by the chord length for $\alpha=0$, and $\alpha=\alpha_c$, as a function of the square root of the logarithm of the chord length.  The expected logarithmic correction is evident for $\alpha=0$, and essentially absent at the transition point: hence the simple 2LR-EPS Ansatz is fully capable of capturing this subtle aspect. Furthermore, a power-law fit of the correlations as a function of the chord length for $\alpha=\alpha_c$ (see inset of Fig.~\ref{Fig:7}) leads to a value of the Luttinger parameter in agreement with the theoretical expectation. Indeed, our estimated value of $K$ is $0.497(5)$.  
 For $\alpha > \alpha_c$, correlations turn to an exponential decay, albeit with an exponentially divergent correlation length for $\alpha \to \alpha_c^+$ \cite{WhiteA1996}, introducing very significant finite-size effects.

 \begin{figure}[b]
\centerline{\includegraphics[width = 1.0\columnwidth]{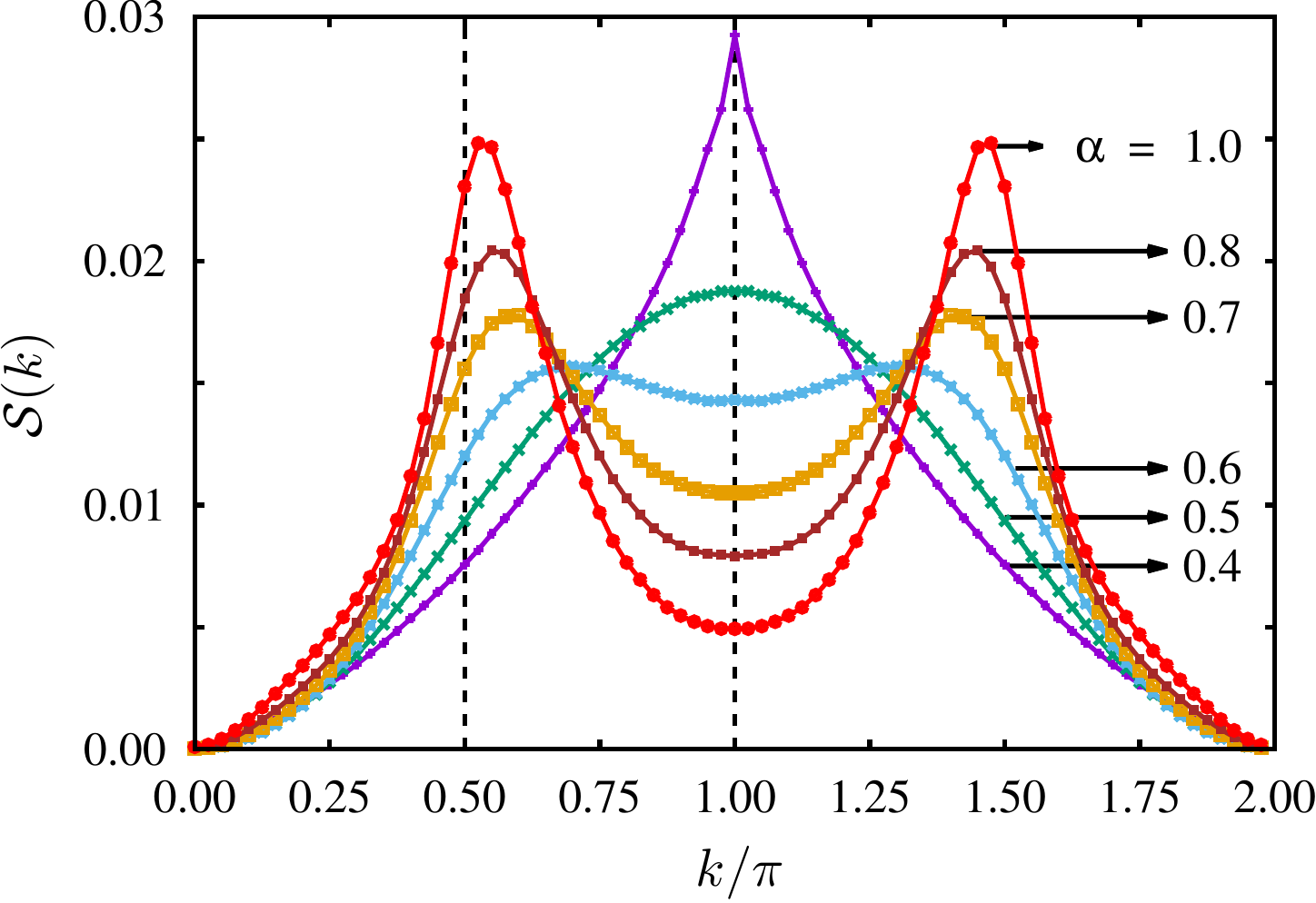}}
\caption{Ground-state spin structure factor for different values of the frustration parameter $\alpha$. Estimates obtained via the $2$LR-EPS variational wave function for a chain with $N=80$. Lines are guides to the eye.} 
\label{Fig:8}
\end{figure}
   \begin{figure}[t]
\centerline{\includegraphics[width = 1.0\columnwidth]{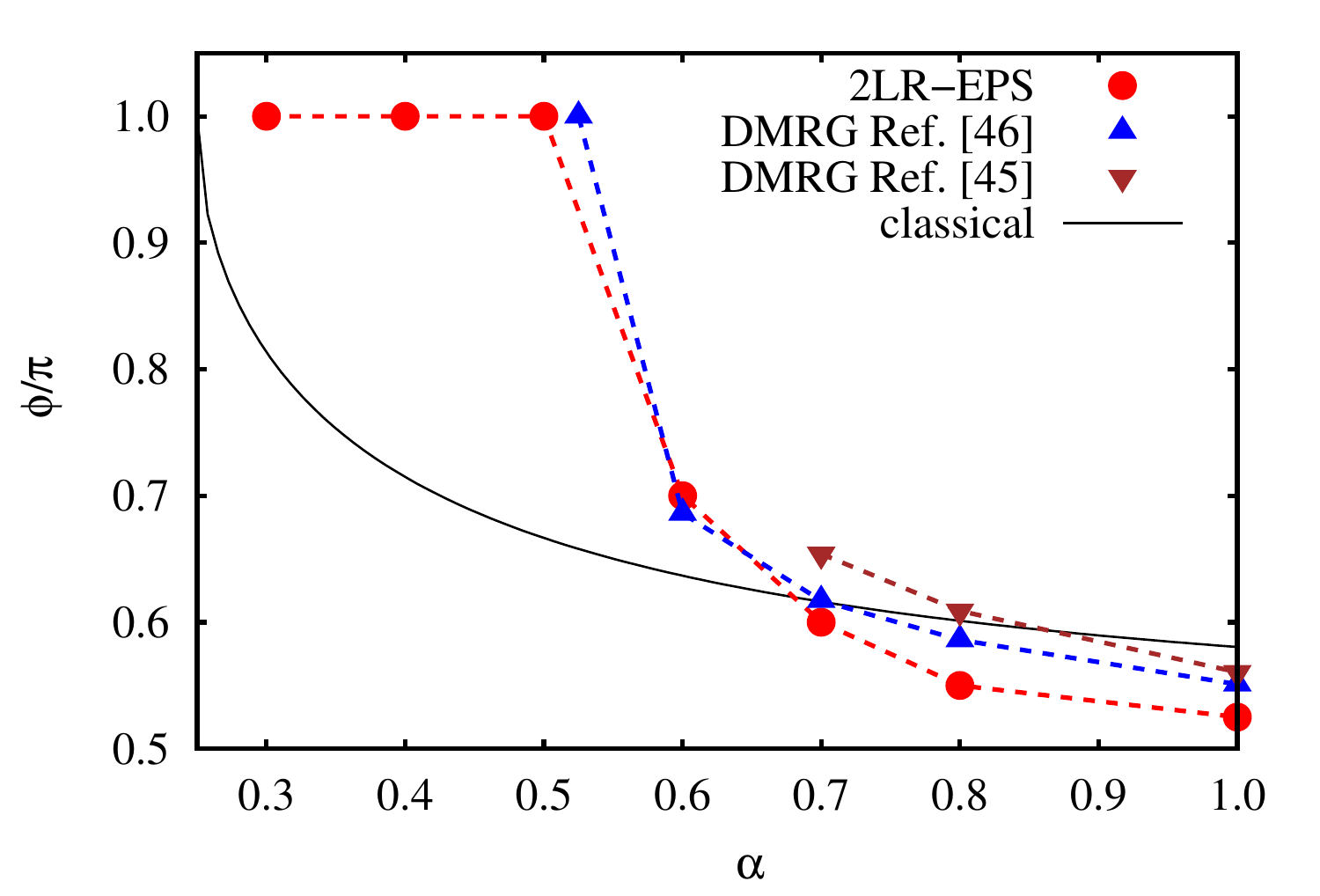}}
\caption{Pitch vector of correlations as a function of $\alpha$.  {Estimates obtained via the $2$LR-EPS variational wave function for a chain with $N=80$ are compared with DMRG results from Refs.~\cite{WhiteA1996,Gehringetal1997} and with the classical result $\phi = \arccos(-1/(4\alpha))$.}} 
\label{Fig:9}
\end{figure}

   \subsubsection{Incommensurability in the correlations for $\alpha > 1/2$} \label{aa}

The incommensurability developing in the correlations for $\alpha > 1/2$ is best seen in the structure factor  ${\cal S}(k) = \sum_\alpha {\cal S}^{\alpha\alpha}(k)$ plotted in Fig.~\ref{Fig:8}. 
There, one clearly observes that the peak at $\pi$, characteristic of N\'eel-like correlations dominant up to $\alpha = 1/2$, splits into two peaks for $\alpha > 1/2$, and the twin peaks move continuously towards the values $\pi/2$ and $3\pi/2$ as $\alpha$ increases. In Fig.~\ref{Fig:9} we compare the pitch vector of correlations $\phi$ namely the position of the left peak in Fig.~\ref{Fig:8}) obtained in this work with previous DMRG results from Refs.~\cite{WhiteA1996, Gehringetal1997} as well as with the MF prediction. The agreement between our estimates and the DMRG ones is acceptable, taking into account that the DMRG results are obtained with open boundary conditions and different system sizes than ours. In general, we consider the successful description of the development of incommensurate helimagnetism at short range as a significant achievement of our Ansatz, considering that it is not accompanied by an obvious breaking of time-reversal invariance. Testing that the phases of all coefficients in the optimized wavefunction are equal modulo $\pi$ is prohibitive when considering the system sizes of interest in this section. Therefore we opt for an alternative test based on spin currents  ${\cal J}_{ij} = i \langle S^+_i S^-_j - S^-_i S^+_j \rangle$, whose average value should be zero on a time-reversal invariant wavefunction, and finite otherwise. Our optimized wavefunctions give a null value of the currents (within the statistical error bar) for all values of $\alpha$, strongly suggesting the real-valuedness of all coefficients. This test is highly non-trivial, as it clearly indicates that the variational optimization exploits the complex nature of coefficients of our Ansatz to introduce incommensurate correlations beyond the reach of a MF-based Ansatz (see Fig.~\ref{Fig:9} and previous discussion in Sec.~\ref{s.J1J2generalities}).
 \begin{figure}[b]
\centerline{\includegraphics[width = 1.0\columnwidth]{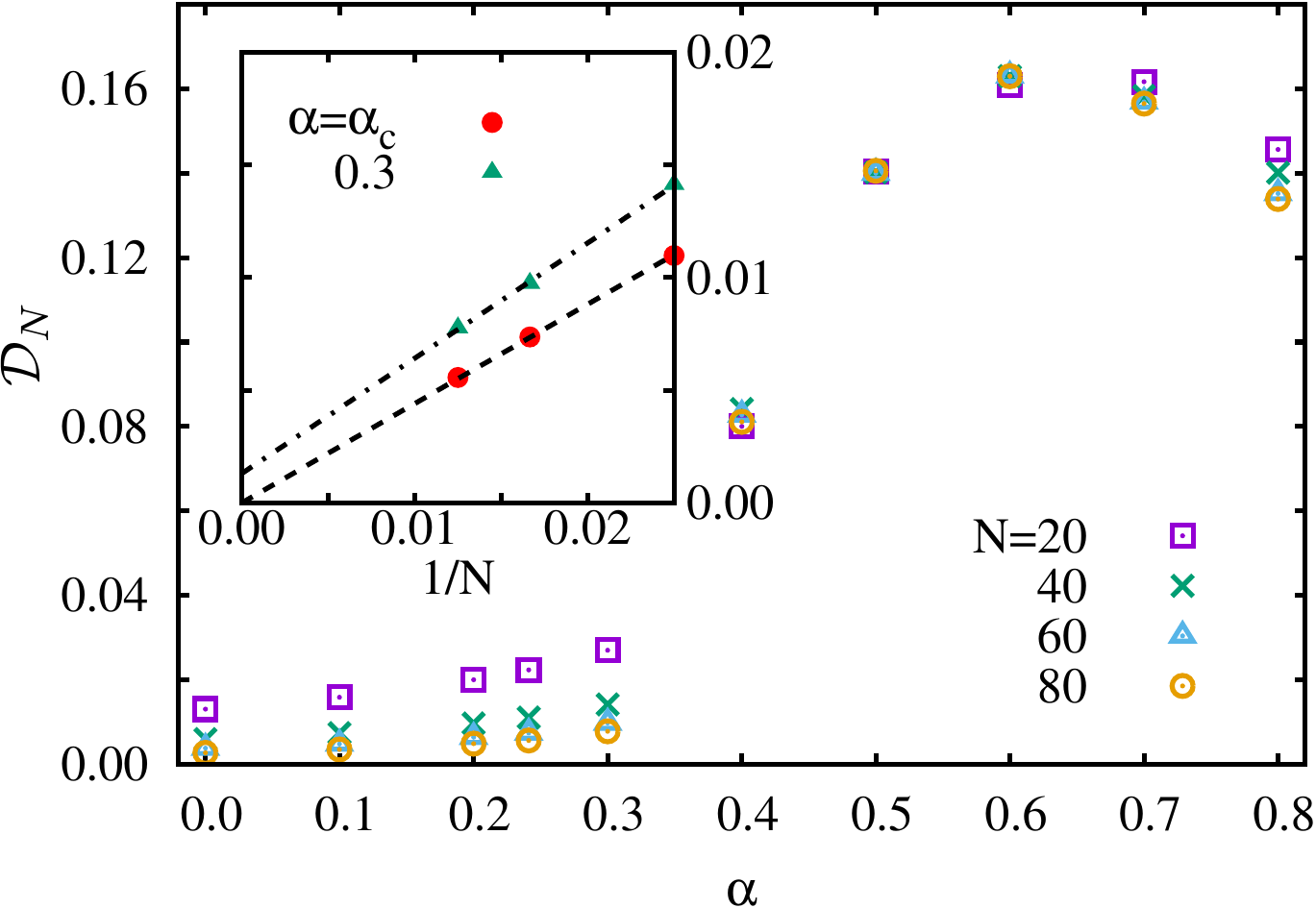}}
\caption{Ground-state dimer order parameter for different system sizes estimated via the $2$LR-EPS Ansatz. Inset: Same quantity as a function of the inverse system size for $\alpha=\alpha_c$ and $\alpha=0.3$ Lines are fits to the corresponding numerical data.} 
\label{Fig:10}
\end{figure}
      
  \subsubsection{VBC order} 
  
 The appearance of VBC order for $\alpha > \alpha_c$ can be efficiently captured by the dimer order parameter $\mathcal D_N$, defined in Eq.~\eqref{e.DOP}.
 
 We show ${\cal D}_N$ for various system sizes as a function of $\alpha$ in Fig.~\ref{Fig:10}, observing that it marks quite clearly the VBC transition. In particular the dimer-dimer correlation function exhibits a decay as $D(r) \sim r^{-1}$ at long distances in the gapless phase, so that we expect that ${\cal D}_N \sim N^{-1}$ in this phase: this is indeed observed in the inset of Fig.~\ref{Fig:10} for $\alpha = \alpha_c$, whereas for $\alpha  = 0.3 > \alpha_c$ one observes that ${\cal D}_N$ extrapolates to a small but finite value.  

\begin{figure}[b]
\centerline{\includegraphics[width = 1.0\columnwidth]{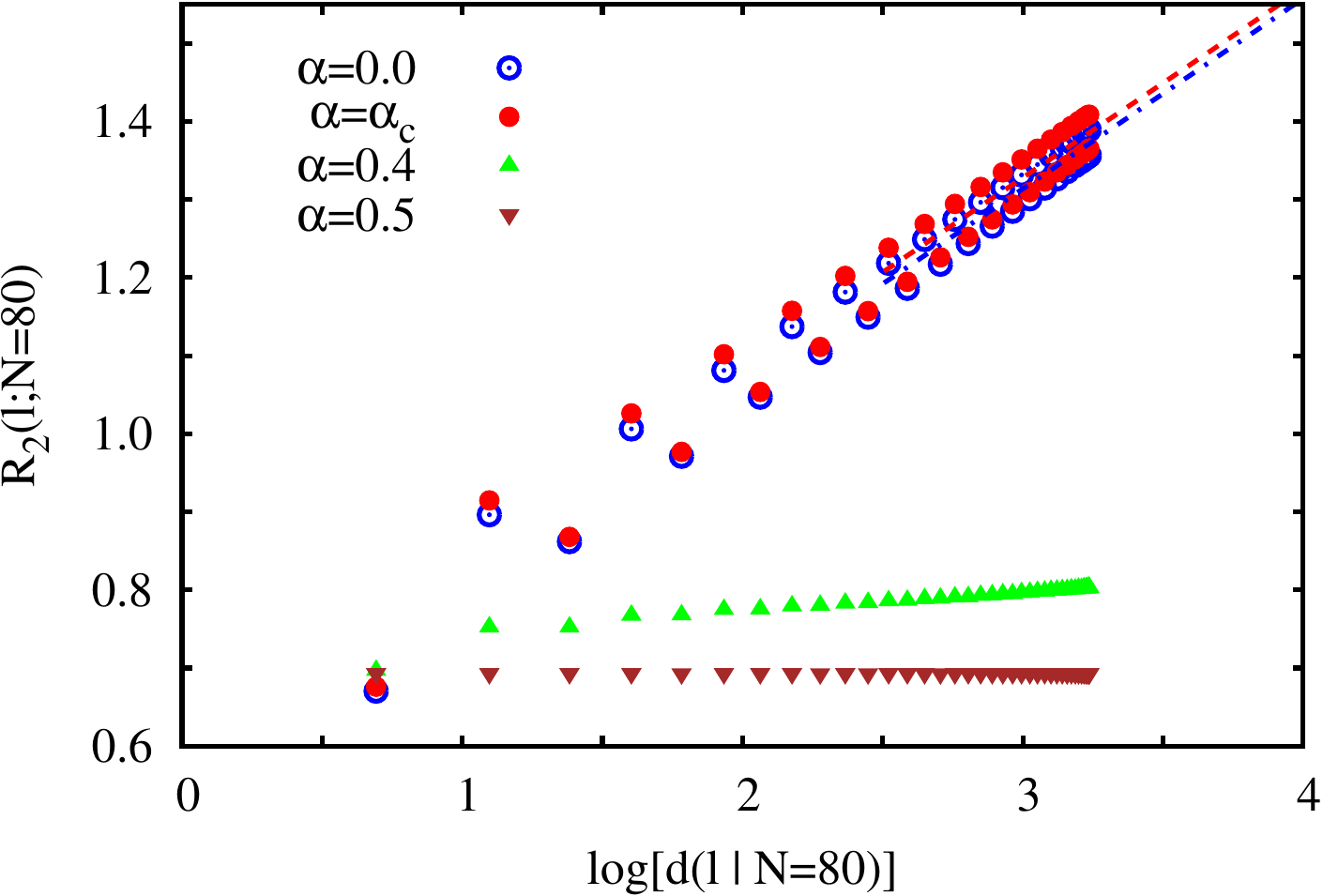}}
\caption{2-R\'enyi entanglement entropy for an $80$-site chain as a function of the logarithm of the chord length $d(l|N)$, $l$ being the subsystem size. The dashed (dotted-dashed) line is a fit to the numerical data with $\alpha=\alpha_c$ and $\alpha=0$. Data shown are obtained with the $2$LR-EPS wave function.} 
\label{Fig:11}
\end{figure}

\begin{figure}[t]
\centerline{\includegraphics[width = 1.0\columnwidth]{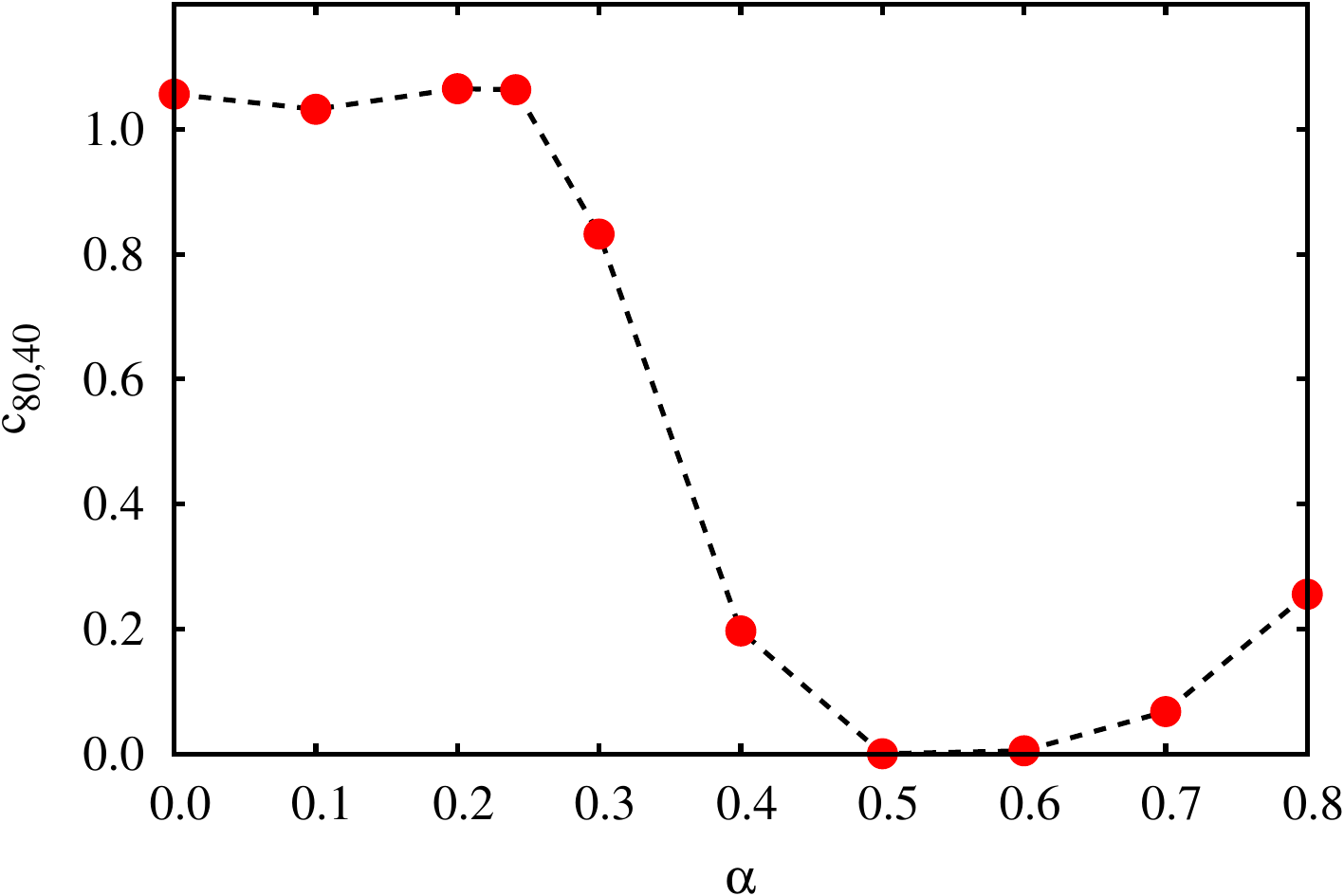}}
\caption{Finite-size central charge estimator defined in the main text as a function of $\alpha$. Estimates are obtained with the $2$LR-EPS wave function.} 
\label{Fig:12}
\end{figure}

  \subsubsection{Scaling of the entanglement entropy}
  
  We conclude our analysis of the 2LR-EPS results for the $J_1-J_2$ chain with a study of the R\'enyi entanglement entropy. Considering a total system of size $N$, in the gapless phase $\alpha < \alpha_c$ the entanglement entropy of a subsystem of linear size $l$ is expected to scale as 
 \begin{equation}
 R_2(l;N) = \frac{c}{4} \log d(l|N) + c_1 + ...
 \label{e.scaling}
 \end{equation} 
where again $c = 1$, and $c_1$ is a non-universal constant \cite{CalabreseC2004} -  extending Eq.~\eqref{e.R2} to account for finite-size effects. On the other hand in the gapped VBC phase the entanglement entropy saturates to a constant (which takes a value $\approx (c/4) \log(\xi)$ close to the critical point), exhibiting therefore an area law. In particular the correlation length is minimal at the MG point, where $R_2(l) \to \log 2$. Fig.~\ref{Fig:11} shows that all these features are very well captured by the 2LR-EPS Ansatz, and in particular the universality of the central charge throughout the gapless phase. 

In order to further contrast the gapless and gapped regime using the entanglement entropy, we concentrate on the difference $R^{N'-N}_2=R_2(N'/2)-R_2(N/2) \simeq (c/4)\log(N'/N)$, which allows one to obtain a finite-size estimate of the central charge as 
 \begin{equation}
 c_{N', N} = \frac{4 R^{N'-N}_2}{\log(N'/N)}
 \end{equation}
  while eliminating the size-independent subleading corrections to the dominant logarithmic scaling (in the gapless phase, see Eq.~\eqref{e.scaling}) \cite{LauchliK2008}.
 Fig.~\ref{Fig:12} shows $c_{80,40}$ as a function of the frustration parameter.  The transition between the gapless and the gapped phase is clearly signalled by the drop of the central charge estimator from values close to unity to vanishing values. On the other hand the estimator increases again for large $\alpha$, consistent with the fact that, in the limit $\alpha \to \infty$, the $J_1-J_2$ model reproduces two decoupled Heisenberg chains with $c=1$.
 
 \section{Conclusions}  
 \label{s.conclusions}
  
  We have discussed a very flexible variational Ansatz (the long-range entangled-plaquette state or LR-EPS) for strongly interacting quantum lattice models, in which explicit quantum correlations within (clusters of) sites are introduced coupling them at all distances into overlapping plaquettes. This Ansatz generalizes both the well-known Jastrow Ansatz (including two-site correlations) \cite{Pang1991} as well as the adjacent-site EPS Ansatz, and  offers the possibility of combining both forms of  overlapping plaquettes  in the same wave function. Moreover, the use of complex coefficients opens the possibility of reproducing non-trivial sign (phase) structures  emerging naturally from the variational optimization without any explicit bias. We have demonstrated the effectiveness of this Ansatz in the case of the frustrated $J_1-J_2$ quantum spin chain: there we show that  it captures both the universal long-wavelength features of the gapless phase of the model (central charge and Luttinger exponent of the corresponding field theory); as well as the appearance of incommensurate short-range correlations at strong frustration. 

 As it currently stands, our Ansatz can be systematically improved by increasing the size of the  plaquettes (either long-range ones or adjacent-site ones) -- this improvement strategy is limited, as it introduces a concomitant exponential growth of the variational parameters. However, our findings show that improving the Ansatz is not crucial to capture the long-range physics, which is already accounted for by using minimal (2-site) long-range plaquettes. Further approaches making the Ansatz systematically improvable with a polynomial growth of the parameters are currently under investigation. Additionally, the success of our wave function in reconstructing non-trivial sign structures variationally suggests its potential application to other models of frustrated quantum magnetism, especially focusing on quantum critical points for which a scale-free variational form is particularly well suited. It would also be very tempting to apply this Ansatz to lattice models of strongly correlated fermions, whose nodal surface becomes the object of variational optimization when using complex coefficients. 
  
\section*{Acknowledgements}
We thank D. Vodola, A. Angelone and G. Misguich for insightful discussions. TR, and FM acknowledge support  from the F\'ed\'eration de Recherche Andr\'e-Marie Amp\`ere (FRAMA). Computer calculations have been performed on the PSMN cluster of the ENS de Lyon.

\appendix

\section{EPS Ansatz and Marshall-like signs}
\label{a.Marshall}

{In the following we show that our simplest Ansatz (namely the 2LR-EPS one) is fully compatible with Marshall-like sign structures, namely sign structures depending uniquely on the parity of the number of $\uparrow$-spins on one of two (arbitrary) sublattices $A$ and $B$ of equal size in which the lattice has been decomposed. 
  A Marshall sign rule with $ABAB...$ sublattice structure is exact in the $J_2\to 0$ limit.  Yet a Marshall-like sign with a sublattice structure  of the kind $ABBA...$ -- or with a longer period -- is also reproducible with the LR-EPS Ansatz; such structures are indeed relevant for the $J_1-J_2$ chain under investigation for sufficiently large $J_2$, as pointed out in \cite{ZengP1995}. It is also important to stress that our Ansatz is compatible with sign structures that are richer than Marshall-like ones.
 
  Marshall-like signs can be represented within the 2LR-EPS Ansatz in several different ways : as the sign/phase of each weight in the wave function results from the sum of the phases of the plaquette coefficients -- similar to the relationship between a vector potential and the flux of the resulting magnetic field -- different parametrizations of Marshall-like signs can be viewed as different gauge choices.
 
The simplest scheme representing Marshall-like signs within a 2LR-EPS amounts to considering that the phase of each plaquette, $\theta_{ij}(\sigma_i,\sigma_j)$, only depends on the configuration of one spin, and it reads
\begin{equation}
\theta_{ij}(\sigma_i,\sigma_j) = \frac{\pi}{N_i} \frac{\sigma_i +1}{2} 
\end{equation} 
if site $i$ belongs to sublattice $A$ or $\theta_{ij}(\sigma_i,\sigma_j) = 0$ otherwise; here and in the following of this Appendix $\sigma_i=\pm 1$ is two times the eigenvalue of the $S_i^z$ operator, and $N_i$ is the number of $(ij)$ plaquettes for which site $i$ is chosen as reference site. This requires therefore the phase of the plaquette to depend on the absolute position of at least one of its sites, leading to ${\sim O}(N)$ parameters. Such a requirement is fully compatible with the choice of modularity of the phases that we made in the study of our largest lattices.

 Alternatively Marshall-like signs can be enforced at the level of each pair: 
 \begin{equation}
 \theta_{ij}(\sigma_i,\sigma_j) = \pi N^{(ij)}_{A,\uparrow} = \frac{\pi}{2} [(\sigma_i+1) \delta_{i,A} + (\sigma_j+1) \delta_{j,A}]  
 \end{equation}
where $N^{(ij)}_{A,\uparrow} = \frac{\pi}{2} [(\sigma_i+1) \delta_{i,A} + (\sigma_j+1) \delta_{j,A}]$ is the number of $\uparrow$ spins on the sites of the $(ij)$ plaquette belonging to sublattice $A$ ($\delta_{i,A} = 1$ if $i$ belongs to $A$ and 0 otherwise). Then one can easily show that, up to a global phase factor: 
\begin{eqnarray}
{\rm sgn}[\psi(\bm \sigma)] & = & e^{ i \frac{\pi}{2} \sum_{i<j} (\sigma_i \delta_{i,A} + \sigma_j \delta_{j,A})} = e^{ i\frac{\pi}{2} (N-1) (2N_{A,\uparrow}-N)} \nonumber \\
& = &  e^{-i\pi N_{A,\uparrow}} e^{i\pi N^2/2} 
\end{eqnarray} 
which is precisely the Marshall sign (up to the constant phase factor $e^{i\pi N^2/2}$). Here we have used the fact that $N$, being composed of two sublattices,  is even, so that $\pi N_{A,\uparrow} N$ is an integer multiple of $2\pi$. We then observe that, in order to reproduce the Marshall-like sign, the phase of the coefficient $e^{i\theta_{ij}}$ must be able to distinguish between $AA$ plaquettes (both sites belonging to $A$) and $BB$ plaquettes (both sides belonging to $B$); and between $AB$ and $BA$ plaquettes. For arbitrary geometries of the $A$ and $B$ sublattices, this requires the phases $\theta_{ij}$ to depend on the absolute positions of \emph{both} sites, increasing substantially the number of parameters to ${\sim O}(N^2)$. We have therefore not pursued this parametrization for the largest lattices studied here.

\section{Emergence of the sign structure and of real wavefunction coefficients during the optimization}
\label{a.sign}
\begin{figure}[t]
\centerline{\includegraphics[width = 1.0\columnwidth]{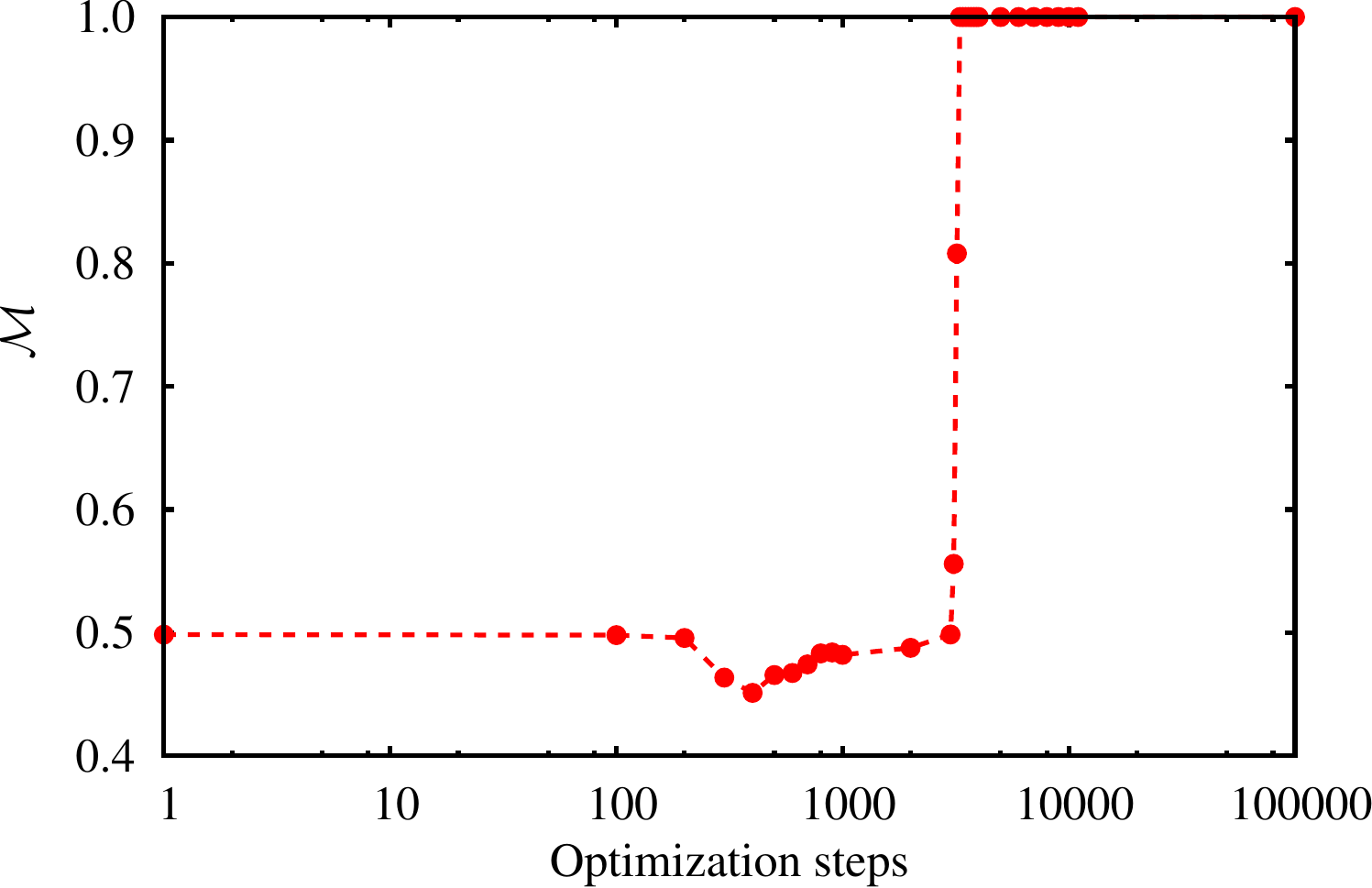}}
\caption{Fraction of the wave function coefficients $\mathcal{M}$ fulfilling the Marshall-sign rule (see text) as a function of the number of optimization steps, for $N=16$, and $\alpha= 0$. Data are obtained with the 2LR-EPS Ansatz. The dashed line is a guide to the eye.} 
\label{Fig:13}
\end{figure}

In order to quantitatively exemplify how the 2LR-EPS wave function is able to reproduce the correct ground state sign structure (when the latter is compatible with the parametrization of the Ansatz) we discuss in detail the case of the $J_1-J_2$ model for $\alpha=0$. Indeed, in this situation, the signs of the coefficients of the exact ground state are known, as they are dictated exactly by the Marshall sign rule,  and they can be exactly reproduced by optimizing our Ansatz.

To elucidate how such a sign structure emerges along the optimization procedure we { closely examine} the evolution with the number of optimization steps of the fraction of wave function coefficients possessing the expected Marshall sign. Hence we define the quantity
\begin{equation} 
\mathcal{M}\!=\!\frac{1}{N_C} \!\!\sum_{\bm \sigma} \frac{1}{2} \Bigg | {\rm sgn}\left [{\rm Re} \left ( \frac{\psi(\bm \sigma)}{\psi(\bm \sigma_{\rm {REF}})} \right)\right ] 
\! + \!\frac{(-1)^{N_{A,\uparrow}(\bm \sigma)}}{(-1)^{N_{A,\uparrow}(\bm \sigma_{\rm {REF}})}}\Bigg | 
\end{equation} 
\begin{figure}[t]
 \vspace{0.5cm}
\centerline{\includegraphics[width = 0.9\columnwidth]{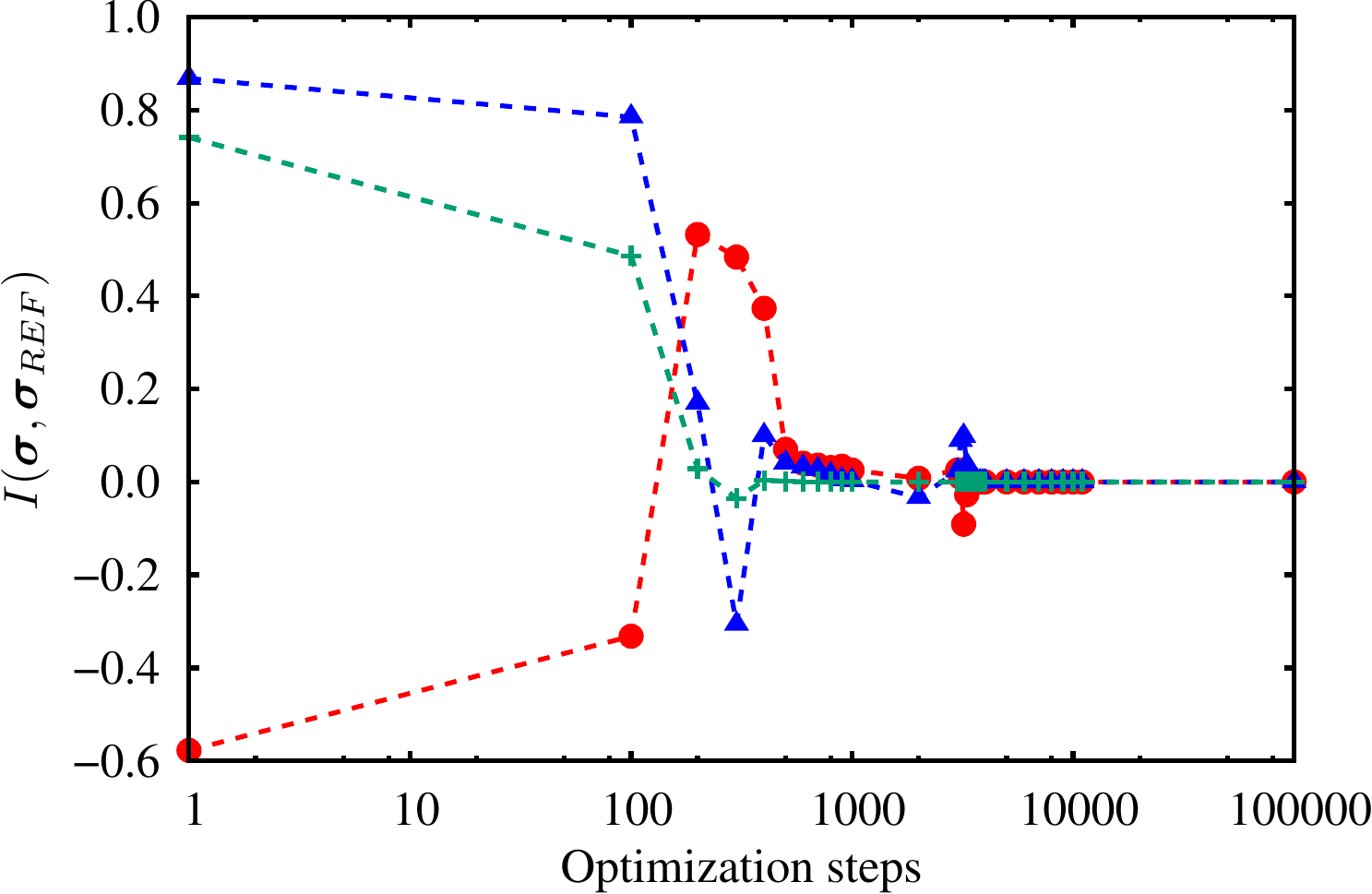}}
\caption{$I(\bm \sigma, \bm \sigma_{\rm {REF}})$ (see text) as a function of the number of optimization steps, for $N=16$, $\alpha= 0$, and 3 (different symbols) randomly chosen wave function coefficients. Data are obtained with the 2LR-EPS Ansatz. Dashed lines are guides to the eye.}
\label{Fig:14}
\end{figure}
where $N_C$ is the total number of { ${\bm \sigma}$} configurations, $\bm \sigma_{\rm {REF}}$ is a reference configuration { (here taken as the N\'eel state $\uparrow \downarrow \uparrow \downarrow ...$)} , and the term in the sum is 1 (0) if the sign of the coefficient of the generic configuration $\bm \sigma$ follows (breaks) the Marshall { sign rule}.

Fig.~\ref{Fig:13} clearly shows how the desired sign structure is fully reconstructed (i.e., $\mathcal{M}$ reaches 1) by our optimization procedure { within a few thousand steps}. 
 
Another important  point of our work is that our optimized 2LR-EPS Ansatz is essentially real (up to a global phase factor), recovering a fundamental property of the ground state of the $J_1-J_2$ chain. 
{ This aspect can be elucidated by analyzing the phase relationship between \emph{e.g.}  the wave-function coefficient of all allowed configurations and that of a 
reference one (here taken again to be the N\'eel state). If the state is real up to a global phase, then the following quantity}
\begin{equation}
 I(\bm \sigma, \bm \sigma_{\rm {REF}})={\rm Im}[\psi(\bm \sigma)/\psi(\bm \sigma_{\rm {REF}})]
 \end{equation}
{ is vanishing}.
{The evolution of $I(\bm \sigma, \bm \sigma_{\rm {REF}})$ along the optimization procedure is shown} in Fig.~\ref{Fig:14} for the same system parameters of Fig.~\ref{Fig:13}, { and} for 3 randomly chosen wave function coefficients. { $I(\bm \sigma, \bm \sigma_{\rm {REF}})$ is seen to approach} a null value (i.e. $< 10^{-15}$) { after a few thousand optimization steps}. { The same behavior can be verified for all the wave function coefficients, and this} despite the complex form of the initial state. 
}
}
\bibliography{paper}

\begin{thebibliography}{52}%
\makeatletter
\providecommand \@ifxundefined [1]{%
 \@ifx{#1\undefined}
}%
\providecommand \@ifnum [1]{%
 \ifnum #1\expandafter \@firstoftwo
 \else \expandafter \@secondoftwo
 \fi
}%
\providecommand \@ifx [1]{%
 \ifx #1\expandafter \@firstoftwo
 \else \expandafter \@secondoftwo
 \fi
}%
\providecommand \natexlab [1]{#1}%
\providecommand \enquote  [1]{``#1''}%
\providecommand \bibnamefont  [1]{#1}%
\providecommand \bibfnamefont [1]{#1}%
\providecommand \citenamefont [1]{#1}%
\providecommand \href@noop [0]{\@secondoftwo}%
\providecommand \href [0]{\begingroup \@sanitize@url \@href}%
\providecommand \@href[1]{\@@startlink{#1}\@@href}%
\providecommand \@@href[1]{\endgroup#1\@@endlink}%
\providecommand \@sanitize@url [0]{\catcode `\\12\catcode `\$12\catcode
  `\&12\catcode `\#12\catcode `\^12\catcode `\_12\catcode `\%12\relax}%
\providecommand \@@startlink[1]{}%
\providecommand \@@endlink[0]{}%
\providecommand \url  [0]{\begingroup\@sanitize@url \@url }%
\providecommand \@url [1]{\endgroup\@href {#1}{\urlprefix }}%
\providecommand \urlprefix  [0]{URL }%
\providecommand \Eprint [0]{\href }%
\providecommand \doibase [0]{http://dx.doi.org/}%
\providecommand \selectlanguage [0]{\@gobble}%
\providecommand \bibinfo  [0]{\@secondoftwo}%
\providecommand \bibfield  [0]{\@secondoftwo}%
\providecommand \translation [1]{[#1]}%
\providecommand \BibitemOpen [0]{}%
\providecommand \bibitemStop [0]{}%
\providecommand \bibitemNoStop [0]{.\EOS\space}%
\providecommand \EOS [0]{\spacefactor3000\relax}%
\providecommand \BibitemShut  [1]{\csname bibitem#1\endcsname}%
\let\auto@bib@innerbib\@empty
\bibitem [{\citenamefont {Sandvik}(2010)}]{Sandvik2010}%
  \BibitemOpen
  \bibfield  {author} {\bibinfo {author} {\bibfnamefont {A.~W.}\ \bibnamefont
  {Sandvik}},\ }\href {\doibase http://dx.doi.org/10.1063/1.3518900} {\bibfield
   {journal} {\bibinfo  {journal} {AIP Conference Proceedings}\ }\textbf
  {\bibinfo {volume} {1297}},\ \bibinfo {pages} {135} (\bibinfo {year}
  {2010})}\BibitemShut {NoStop}%
\bibitem [{\citenamefont {Becca}\ and\ \citenamefont
  {Sorella}(2017)}]{BeccaS2017}%
  \BibitemOpen
  \bibfield  {author} {\bibinfo {author} {\bibfnamefont {F.}~\bibnamefont
  {Becca}}\ and\ \bibinfo {author} {\bibfnamefont {S.}~\bibnamefont
  {Sorella}},\ }\href@noop {} {\emph {\bibinfo {title} {Quantum Monte Carlo
  Approaches for Correlated Systems}}}\ (\bibinfo  {publisher} {Cambridge},\
  \bibinfo {year} {2017})\BibitemShut {NoStop}%
\bibitem [{\citenamefont {Chandrasekharan}\ and\ \citenamefont
  {Wiese}(1999)}]{ChandrasekharanW1999}%
  \BibitemOpen
  \bibfield  {author} {\bibinfo {author} {\bibfnamefont {S.}~\bibnamefont
  {Chandrasekharan}}\ and\ \bibinfo {author} {\bibfnamefont {U.-J.}\
  \bibnamefont {Wiese}},\ }\href {\doibase 10.1103/PhysRevLett.83.3116}
  {\bibfield  {journal} {\bibinfo  {journal} {Phys. Rev. Lett.}\ }\textbf
  {\bibinfo {volume} {83}},\ \bibinfo {pages} {3116} (\bibinfo {year}
  {1999})}\BibitemShut {NoStop}%
\bibitem [{\citenamefont {Schollw\"ock}(2005)}]{Schollwoeck2005}%
  \BibitemOpen
  \bibfield  {author} {\bibinfo {author} {\bibfnamefont {U.}~\bibnamefont
  {Schollw\"ock}},\ }\href {\doibase 10.1103/RevModPhys.77.259} {\bibfield
  {journal} {\bibinfo  {journal} {Rev. Mod. Phys.}\ }\textbf {\bibinfo {volume}
  {77}},\ \bibinfo {pages} {259} (\bibinfo {year} {2005})}\BibitemShut
  {NoStop}%
\bibitem [{\citenamefont {Schollw\"ock}(2011)}]{Schollwoeck2011}%
  \BibitemOpen
  \bibfield  {author} {\bibinfo {author} {\bibfnamefont {U.}~\bibnamefont
  {Schollw\"ock}},\ }\href {\doibase https://doi.org/10.1016/j.aop.2010.09.012}
  {\bibfield  {journal} {\bibinfo  {journal} {Annals of Physics}\ }\textbf
  {\bibinfo {volume} {326}},\ \bibinfo {pages} {96 } (\bibinfo {year}
  {2011})}\BibitemShut {NoStop}%
\bibitem [{\citenamefont {Or\'us}(2014)}]{Orus2014}%
  \BibitemOpen
  \bibfield  {author} {\bibinfo {author} {\bibfnamefont {R.}~\bibnamefont
  {Or\'us}},\ }\href {\doibase https://doi.org/10.1016/j.aop.2014.06.013}
  {\bibfield  {journal} {\bibinfo  {journal} {Annals of Physics}\ }\textbf
  {\bibinfo {volume} {349}},\ \bibinfo {pages} {117 } (\bibinfo {year}
  {2014})}\BibitemShut {NoStop}%
\bibitem [{\citenamefont {Verstraete}\ \emph {et~al.}(2008)\citenamefont
  {Verstraete}, \citenamefont {Murg},\ and\ \citenamefont
  {Cirac}}]{Verstraeteetal2008}%
  \BibitemOpen
  \bibfield  {author} {\bibinfo {author} {\bibfnamefont {F.}~\bibnamefont
  {Verstraete}}, \bibinfo {author} {\bibfnamefont {V.}~\bibnamefont {Murg}}, \
  and\ \bibinfo {author} {\bibfnamefont {J.}~\bibnamefont {Cirac}},\ }\href
  {\doibase 10.1080/14789940801912366} {\bibfield  {journal} {\bibinfo
  {journal} {Advances in Physics}\ }\textbf {\bibinfo {volume} {57}},\ \bibinfo
  {pages} {143} (\bibinfo {year} {2008})},\ \Eprint
  {http://arxiv.org/abs/https://doi.org/10.1080/14789940801912366}
  {https://doi.org/10.1080/14789940801912366} \BibitemShut {NoStop}%
\bibitem [{\citenamefont {Liang}\ \emph {et~al.}(1988)\citenamefont {Liang},
  \citenamefont {Doucot},\ and\ \citenamefont {Anderson}}]{Liangetal1988}%
  \BibitemOpen
  \bibfield  {author} {\bibinfo {author} {\bibfnamefont {S.}~\bibnamefont
  {Liang}}, \bibinfo {author} {\bibfnamefont {B.}~\bibnamefont {Doucot}}, \
  and\ \bibinfo {author} {\bibfnamefont {P.~W.}\ \bibnamefont {Anderson}},\
  }\href {\doibase 10.1103/PhysRevLett.61.365} {\bibfield  {journal} {\bibinfo
  {journal} {Phys. Rev. Lett.}\ }\textbf {\bibinfo {volume} {61}},\ \bibinfo
  {pages} {365} (\bibinfo {year} {1988})}\BibitemShut {NoStop}%
\bibitem [{\citenamefont {Becca}\ \emph {et~al.}(2010)\citenamefont {Becca},
  \citenamefont {Capriotti}, \citenamefont {Parola},\ and\ \citenamefont
  {Sorella}}]{Beccaetal2010}%
  \BibitemOpen
  \bibfield  {author} {\bibinfo {author} {\bibfnamefont {F.}~\bibnamefont
  {Becca}}, \bibinfo {author} {\bibfnamefont {L.}~\bibnamefont {Capriotti}},
  \bibinfo {author} {\bibfnamefont {A.}~\bibnamefont {Parola}}, \ and\ \bibinfo
  {author} {\bibfnamefont {S.}~\bibnamefont {Sorella}},\ }in\ \href@noop {}
  {\emph {\bibinfo {booktitle} {Introduction to Frustrated Magnetism}}},\
  \bibinfo {editor} {edited by\ \bibinfo {editor} {\bibfnamefont
  {C.}~\bibnamefont {Lacroix}}, \bibinfo {editor} {\bibfnamefont
  {P.}~\bibnamefont {Mendels}}, \ and\ \bibinfo {editor} {\bibfnamefont
  {F.}~\bibnamefont {Mila}}}\ (\bibinfo  {publisher} {Springer},\ \bibinfo
  {year} {2010})\BibitemShut {NoStop}%
\bibitem [{\citenamefont {Carleo}\ and\ \citenamefont
  {Troyer}(2017)}]{CarleoT2017}%
  \BibitemOpen
  \bibfield  {author} {\bibinfo {author} {\bibfnamefont {G.}~\bibnamefont
  {Carleo}}\ and\ \bibinfo {author} {\bibfnamefont {M.}~\bibnamefont
  {Troyer}},\ }\href {\doibase 10.1126/science.aag2302} {\bibfield  {journal}
  {\bibinfo  {journal} {Science}\ }\textbf {\bibinfo {volume} {355}},\ \bibinfo
  {pages} {602} (\bibinfo {year} {2017})},\ \Eprint
  {http://arxiv.org/abs/http://science.sciencemag.org/content/355/6325/602.full.pdf}
  {http://science.sciencemag.org/content/355/6325/602.full.pdf} \BibitemShut
  {NoStop}%
\bibitem [{\citenamefont {Mezzacapo}\ \emph {et~al.}(2009)\citenamefont
  {Mezzacapo}, \citenamefont {Schuch}, \citenamefont {Boninsegni},\ and\
  \citenamefont {Cirac}}]{Mezzacapoetal2009}%
  \BibitemOpen
  \bibfield  {author} {\bibinfo {author} {\bibfnamefont {F.}~\bibnamefont
  {Mezzacapo}}, \bibinfo {author} {\bibfnamefont {N.}~\bibnamefont {Schuch}},
  \bibinfo {author} {\bibfnamefont {M.}~\bibnamefont {Boninsegni}}, \ and\
  \bibinfo {author} {\bibfnamefont {J.~I.}\ \bibnamefont {Cirac}},\ }\href
  {\doibase 10.1088/1367-2630/11/8/083026} {\bibfield  {journal} {\bibinfo
  {journal} {New Journal of Physics}\ }\textbf {\bibinfo {volume} {11}},\
  \bibinfo {pages} {083026} (\bibinfo {year} {2009})}\BibitemShut {NoStop}%
\bibitem [{\citenamefont {Changlani}\ \emph {et~al.}(2009)\citenamefont
  {Changlani}, \citenamefont {Kinder}, \citenamefont {Umrigar},\ and\
  \citenamefont {Chan}}]{Changlanietal2009}%
  \BibitemOpen
  \bibfield  {author} {\bibinfo {author} {\bibfnamefont {H.~J.}\ \bibnamefont
  {Changlani}}, \bibinfo {author} {\bibfnamefont {J.~M.}\ \bibnamefont
  {Kinder}}, \bibinfo {author} {\bibfnamefont {C.~J.}\ \bibnamefont {Umrigar}},
  \ and\ \bibinfo {author} {\bibfnamefont {G.~K.-L.}\ \bibnamefont {Chan}},\
  }\href {\doibase 10.1103/PhysRevB.80.245116} {\bibfield  {journal} {\bibinfo
  {journal} {Phys. Rev. B}\ }\textbf {\bibinfo {volume} {80}},\ \bibinfo
  {pages} {245116} (\bibinfo {year} {2009})}\BibitemShut {NoStop}%
\bibitem [{\citenamefont {Gao}\ and\ \citenamefont {Duan}(2017)}]{GaoD2017}%
  \BibitemOpen
  \bibfield  {author} {\bibinfo {author} {\bibfnamefont {X.}~\bibnamefont
  {Gao}}\ and\ \bibinfo {author} {\bibfnamefont {L.-M.}\ \bibnamefont {Duan}},\
  }\href {\doibase 10.1038/s41467-017-00705-2} {\bibfield  {journal} {\bibinfo
  {journal} {Nature Communications}\ }\textbf {\bibinfo {volume} {8}},\
  \bibinfo {pages} {662} (\bibinfo {year} {2017})}\BibitemShut {NoStop}%
\bibitem [{\citenamefont {{Choo}}\ \emph {et~al.}(2019)\citenamefont {{Choo}},
  \citenamefont {{Neupert}},\ and\ \citenamefont {{Carleo}}}]{Chooetal2019}%
  \BibitemOpen
  \bibfield  {author} {\bibinfo {author} {\bibfnamefont {K.}~\bibnamefont
  {{Choo}}}, \bibinfo {author} {\bibfnamefont {T.}~\bibnamefont {{Neupert}}}, \
  and\ \bibinfo {author} {\bibfnamefont {G.}~\bibnamefont {{Carleo}}},\
  }\href@noop {} {\bibfield  {journal} {\bibinfo  {journal} {arXiv e-prints}\
  ,\ \bibinfo {eid} {arXiv:1903.06713}} (\bibinfo {year} {2019})},\ \Eprint
  {http://arxiv.org/abs/1903.06713} {arXiv:1903.06713 [cond-mat.str-el]}
  \BibitemShut {NoStop}%
\bibitem [{\citenamefont {Cai}\ and\ \citenamefont {Liu}(2018)}]{Caietal2018}%
  \BibitemOpen
  \bibfield  {author} {\bibinfo {author} {\bibfnamefont {Z.}~\bibnamefont
  {Cai}}\ and\ \bibinfo {author} {\bibfnamefont {J.}~\bibnamefont {Liu}},\
  }\href {\doibase 10.1103/PhysRevB.97.035116} {\bibfield  {journal} {\bibinfo
  {journal} {Phys. Rev. B}\ }\textbf {\bibinfo {volume} {97}},\ \bibinfo
  {pages} {035116} (\bibinfo {year} {2018})}\BibitemShut {NoStop}%
\bibitem [{\citenamefont {Mezzacapo}\ and\ \citenamefont
  {Cirac}(2010)}]{MezzacapoC2010}%
  \BibitemOpen
  \bibfield  {author} {\bibinfo {author} {\bibfnamefont {F.}~\bibnamefont
  {Mezzacapo}}\ and\ \bibinfo {author} {\bibfnamefont {J.~I.}\ \bibnamefont
  {Cirac}},\ }\href {\doibase 10.1088/1367-2630/12/10/103039} {\bibfield
  {journal} {\bibinfo  {journal} {New Journal of Physics}\ }\textbf {\bibinfo
  {volume} {12}},\ \bibinfo {pages} {103039} (\bibinfo {year}
  {2010})}\BibitemShut {NoStop}%
\bibitem [{\citenamefont {Mezzacapo}(2011)}]{Mezzacapo2011}%
  \BibitemOpen
  \bibfield  {author} {\bibinfo {author} {\bibfnamefont {F.}~\bibnamefont
  {Mezzacapo}},\ }\href {\doibase 10.1103/PhysRevB.83.115111} {\bibfield
  {journal} {\bibinfo  {journal} {Phys. Rev. B}\ }\textbf {\bibinfo {volume}
  {83}},\ \bibinfo {pages} {115111} (\bibinfo {year} {2011})}\BibitemShut
  {NoStop}%
\bibitem [{\citenamefont {Al-Assam}\ \emph {et~al.}(2011)\citenamefont
  {Al-Assam}, \citenamefont {Clark}, \citenamefont {Foot},\ and\ \citenamefont
  {Jaksch}}]{Al-Assametal2011}%
  \BibitemOpen
  \bibfield  {author} {\bibinfo {author} {\bibfnamefont {S.}~\bibnamefont
  {Al-Assam}}, \bibinfo {author} {\bibfnamefont {S.~R.}\ \bibnamefont {Clark}},
  \bibinfo {author} {\bibfnamefont {C.~J.}\ \bibnamefont {Foot}}, \ and\
  \bibinfo {author} {\bibfnamefont {D.}~\bibnamefont {Jaksch}},\ }\href
  {\doibase 10.1103/PhysRevB.84.205108} {\bibfield  {journal} {\bibinfo
  {journal} {Phys. Rev. B}\ }\textbf {\bibinfo {volume} {84}},\ \bibinfo
  {pages} {205108} (\bibinfo {year} {2011})}\BibitemShut {NoStop}%
\bibitem [{\citenamefont {Mezzacapo}\ and\ \citenamefont
  {Boninsegni}(2012)}]{MezzacapoB2012}%
  \BibitemOpen
  \bibfield  {author} {\bibinfo {author} {\bibfnamefont {F.}~\bibnamefont
  {Mezzacapo}}\ and\ \bibinfo {author} {\bibfnamefont {M.}~\bibnamefont
  {Boninsegni}},\ }\href {\doibase 10.1103/PhysRevB.85.060402} {\bibfield
  {journal} {\bibinfo  {journal} {Phys. Rev. B}\ }\textbf {\bibinfo {volume}
  {85}},\ \bibinfo {pages} {060402} (\bibinfo {year} {2012})}\BibitemShut
  {NoStop}%
\bibitem [{\citenamefont {Mezzacapo}(2012)}]{Mezzacapo2012}%
  \BibitemOpen
  \bibfield  {author} {\bibinfo {author} {\bibfnamefont {F.}~\bibnamefont
  {Mezzacapo}},\ }\href {\doibase 10.1103/PhysRevB.86.045115} {\bibfield
  {journal} {\bibinfo  {journal} {Phys. Rev. B}\ }\textbf {\bibinfo {volume}
  {86}},\ \bibinfo {pages} {045115} (\bibinfo {year} {2012})}\BibitemShut
  {NoStop}%
\bibitem [{\citenamefont {Duri\v{c}}\ \emph {et~al.}(2014)\citenamefont
  {Duri\v{c}}, \citenamefont {Chancellor},\ and\ \citenamefont
  {Herbut}}]{Duricetal2014}%
  \BibitemOpen
  \bibfield  {author} {\bibinfo {author} {\bibfnamefont {T.}~\bibnamefont
  {Duri\v{c}}}, \bibinfo {author} {\bibfnamefont {N.}~\bibnamefont
  {Chancellor}}, \ and\ \bibinfo {author} {\bibfnamefont {I.~F.}\ \bibnamefont
  {Herbut}},\ }\href {\doibase 10.1103/PhysRevB.89.165123} {\bibfield
  {journal} {\bibinfo  {journal} {Phys. Rev. B}\ }\textbf {\bibinfo {volume}
  {89}},\ \bibinfo {pages} {165123} (\bibinfo {year} {2014})}\BibitemShut
  {NoStop}%
\bibitem [{\citenamefont {Duri\v{c}}\ \emph {et~al.}(2016)\citenamefont
  {Duri\v{c}}, \citenamefont {Chancellor}, \citenamefont {Crowley},
  \citenamefont {Di~Cintio},\ and\ \citenamefont {Green}}]{Duricetal2016}%
  \BibitemOpen
  \bibfield  {author} {\bibinfo {author} {\bibfnamefont {T.}~\bibnamefont
  {Duri\v{c}}}, \bibinfo {author} {\bibfnamefont {N.}~\bibnamefont
  {Chancellor}}, \bibinfo {author} {\bibfnamefont {P.~J.~D.}\ \bibnamefont
  {Crowley}}, \bibinfo {author} {\bibfnamefont {P.}~\bibnamefont {Di~Cintio}},
  \ and\ \bibinfo {author} {\bibfnamefont {A.~G.}\ \bibnamefont {Green}},\
  }\href {\doibase 10.1103/PhysRevB.93.085143} {\bibfield  {journal} {\bibinfo
  {journal} {Phys. Rev. B}\ }\textbf {\bibinfo {volume} {93}},\ \bibinfo
  {pages} {085143} (\bibinfo {year} {2016})}\BibitemShut {NoStop}%
\bibitem [{\citenamefont {Mezzacapo}\ \emph {et~al.}(2016)\citenamefont
  {Mezzacapo}, \citenamefont {Angelone},\ and\ \citenamefont
  {Pupillo}}]{Mezzacapoetal2016}%
  \BibitemOpen
  \bibfield  {author} {\bibinfo {author} {\bibfnamefont {F.}~\bibnamefont
  {Mezzacapo}}, \bibinfo {author} {\bibfnamefont {A.}~\bibnamefont {Angelone}},
  \ and\ \bibinfo {author} {\bibfnamefont {G.}~\bibnamefont {Pupillo}},\ }\href
  {\doibase 10.1103/PhysRevB.94.155120} {\bibfield  {journal} {\bibinfo
  {journal} {Phys. Rev. B}\ }\textbf {\bibinfo {volume} {94}},\ \bibinfo
  {pages} {155120} (\bibinfo {year} {2016})}\BibitemShut {NoStop}%
\bibitem [{\citenamefont {Sutherland}(2004)}]{Sutherland-book}%
  \BibitemOpen
  \bibfield  {author} {\bibinfo {author} {\bibfnamefont {B.}~\bibnamefont
  {Sutherland}},\ }\href@noop {} {\emph {\bibinfo {title} {Beautiful Models: 70
  Years of Exactly Solved Quantum Many-body Problems}}}\ (\bibinfo  {publisher}
  {World Scientific},\ \bibinfo {address} {Singapore},\ \bibinfo {year}
  {2004})\BibitemShut {NoStop}%
\bibitem [{\citenamefont {Stojevic}\ \emph {et~al.}(2016)\citenamefont
  {Stojevic}, \citenamefont {Crowley}, \citenamefont {Duri\v{c}}, \citenamefont
  {Grey},\ and\ \citenamefont {Green}}]{Stojevicetal2016}%
  \BibitemOpen
  \bibfield  {author} {\bibinfo {author} {\bibfnamefont {V.}~\bibnamefont
  {Stojevic}}, \bibinfo {author} {\bibfnamefont {P.}~\bibnamefont {Crowley}},
  \bibinfo {author} {\bibfnamefont {T.}~\bibnamefont {Duri\v{c}}}, \bibinfo
  {author} {\bibfnamefont {C.}~\bibnamefont {Grey}}, \ and\ \bibinfo {author}
  {\bibfnamefont {A.~G.}\ \bibnamefont {Green}},\ }\href {\doibase
  10.1103/PhysRevB.94.165135} {\bibfield  {journal} {\bibinfo  {journal} {Phys.
  Rev. B}\ }\textbf {\bibinfo {volume} {94}},\ \bibinfo {pages} {165135}
  (\bibinfo {year} {2016})}\BibitemShut {NoStop}%
\bibitem [{\citenamefont {Pang}(1991)}]{Pang1991}%
  \BibitemOpen
  \bibfield  {author} {\bibinfo {author} {\bibfnamefont {T.}~\bibnamefont
  {Pang}},\ }\href {\doibase 10.1103/PhysRevB.43.3362} {\bibfield  {journal}
  {\bibinfo  {journal} {Phys. Rev. B}\ }\textbf {\bibinfo {volume} {43}},\
  \bibinfo {pages} {3362} (\bibinfo {year} {1991})}\BibitemShut {NoStop}%
\bibitem [{\citenamefont {Kramer}\ and\ \citenamefont
  {Saraceno}(1981)}]{bookTDVP}%
  \BibitemOpen
  \bibfield  {author} {\bibinfo {author} {\bibfnamefont {P.}~\bibnamefont
  {Kramer}}\ and\ \bibinfo {author} {\bibfnamefont {M.}~\bibnamefont
  {Saraceno}},\ }\href@noop {} {\emph {\bibinfo {title} {Geometry of the
  Time-Dependent Variational Principle in Quantum Mechanics}}}\ (\bibinfo
  {publisher} {Springer},\ \bibinfo {year} {1981})\BibitemShut {NoStop}%
\bibitem [{\citenamefont {Auerbach}(2006)}]{Auerbachbook}%
  \BibitemOpen
  \bibfield  {author} {\bibinfo {author} {\bibfnamefont {A.}~\bibnamefont
  {Auerbach}},\ }\href@noop {} {\emph {\bibinfo {title} {Interacting Electrons
  and Quantum Magnetism}}}\ (\bibinfo  {publisher} {Springer},\ \bibinfo {year}
  {2006})\BibitemShut {NoStop}%
\bibitem [{\citenamefont {Hastings}\ \emph {et~al.}(2010)\citenamefont
  {Hastings}, \citenamefont {Gonz\'alez}, \citenamefont {Kallin},\ and\
  \citenamefont {Melko}}]{Hastingsetal2010}%
  \BibitemOpen
  \bibfield  {author} {\bibinfo {author} {\bibfnamefont {M.~B.}\ \bibnamefont
  {Hastings}}, \bibinfo {author} {\bibfnamefont {I.}~\bibnamefont
  {Gonz\'alez}}, \bibinfo {author} {\bibfnamefont {A.~B.}\ \bibnamefont
  {Kallin}}, \ and\ \bibinfo {author} {\bibfnamefont {R.~G.}\ \bibnamefont
  {Melko}},\ }\href {\doibase 10.1103/PhysRevLett.104.157201} {\bibfield
  {journal} {\bibinfo  {journal} {Phys. Rev. Lett.}\ }\textbf {\bibinfo
  {volume} {104}},\ \bibinfo {pages} {157201} (\bibinfo {year}
  {2010})}\BibitemShut {NoStop}%
\bibitem [{\citenamefont {Stoudenmire}\ and\ \citenamefont
  {White}(2012)}]{StoudenmireW2012}%
  \BibitemOpen
  \bibfield  {author} {\bibinfo {author} {\bibfnamefont {E.}~\bibnamefont
  {Stoudenmire}}\ and\ \bibinfo {author} {\bibfnamefont {S.~R.}\ \bibnamefont
  {White}},\ }\href {\doibase 10.1146/annurev-conmatphys-020911-125018}
  {\bibfield  {journal} {\bibinfo  {journal} {Annual Review of Condensed Matter
  Physics}\ }\textbf {\bibinfo {volume} {3}},\ \bibinfo {pages} {111} (\bibinfo
  {year} {2012})},\ \Eprint
  {http://arxiv.org/abs/https://doi.org/10.1146/annurev-conmatphys-020911-125018}
  {https://doi.org/10.1146/annurev-conmatphys-020911-125018} \BibitemShut
  {NoStop}%
\bibitem [{\citenamefont {Corboz}\ \emph {et~al.}(2018)\citenamefont {Corboz},
  \citenamefont {Czarnik}, \citenamefont {Kapteijns},\ and\ \citenamefont
  {Tagliacozzo}}]{Corbozetal2018}%
  \BibitemOpen
  \bibfield  {author} {\bibinfo {author} {\bibfnamefont {P.}~\bibnamefont
  {Corboz}}, \bibinfo {author} {\bibfnamefont {P.}~\bibnamefont {Czarnik}},
  \bibinfo {author} {\bibfnamefont {G.}~\bibnamefont {Kapteijns}}, \ and\
  \bibinfo {author} {\bibfnamefont {L.}~\bibnamefont {Tagliacozzo}},\ }\href
  {\doibase 10.1103/PhysRevX.8.031031} {\bibfield  {journal} {\bibinfo
  {journal} {Phys. Rev. X}\ }\textbf {\bibinfo {volume} {8}},\ \bibinfo {pages}
  {031031} (\bibinfo {year} {2018})}\BibitemShut {NoStop}%
\bibitem [{\citenamefont {Glasser}\ \emph {et~al.}(2018)\citenamefont
  {Glasser}, \citenamefont {Pancotti}, \citenamefont {August}, \citenamefont
  {Rodriguez},\ and\ \citenamefont {Cirac}}]{Glasseretal2018}%
  \BibitemOpen
  \bibfield  {author} {\bibinfo {author} {\bibfnamefont {I.}~\bibnamefont
  {Glasser}}, \bibinfo {author} {\bibfnamefont {N.}~\bibnamefont {Pancotti}},
  \bibinfo {author} {\bibfnamefont {M.}~\bibnamefont {August}}, \bibinfo
  {author} {\bibfnamefont {I.~D.}\ \bibnamefont {Rodriguez}}, \ and\ \bibinfo
  {author} {\bibfnamefont {J.~I.}\ \bibnamefont {Cirac}},\ }\href {\doibase
  10.1103/PhysRevX.8.011006} {\bibfield  {journal} {\bibinfo  {journal} {Phys.
  Rev. X}\ }\textbf {\bibinfo {volume} {8}},\ \bibinfo {pages} {011006}
  (\bibinfo {year} {2018})}\BibitemShut {NoStop}%
\bibitem [{\citenamefont {Anderson}(1987)}]{Anderson1987}%
  \BibitemOpen
  \bibfield  {author} {\bibinfo {author} {\bibfnamefont {P.~W.}\ \bibnamefont
  {Anderson}},\ }\href {\doibase 10.1126/science.235.4793.1196} {\bibfield
  {journal} {\bibinfo  {journal} {Science}\ }\textbf {\bibinfo {volume}
  {235}},\ \bibinfo {pages} {1196} (\bibinfo {year} {1987})}\BibitemShut
  {NoStop}%
\bibitem [{\citenamefont {Lieb}\ \emph {et~al.}(1961)\citenamefont {Lieb},
  \citenamefont {Schultz},\ and\ \citenamefont {Mattis}}]{Liebetal1961}%
  \BibitemOpen
  \bibfield  {author} {\bibinfo {author} {\bibfnamefont {E.}~\bibnamefont
  {Lieb}}, \bibinfo {author} {\bibfnamefont {T.}~\bibnamefont {Schultz}}, \
  and\ \bibinfo {author} {\bibfnamefont {D.}~\bibnamefont {Mattis}},\ }\href
  {\doibase https://doi.org/10.1016/0003-4916(61)90115-4} {\bibfield  {journal}
  {\bibinfo  {journal} {Annals of Physics}\ }\textbf {\bibinfo {volume} {16}},\
  \bibinfo {pages} {407 } (\bibinfo {year} {1961})}\BibitemShut {NoStop}%
\bibitem [{\citenamefont {Sutherland}(1971)}]{Sutherland1971}%
  \BibitemOpen
  \bibfield  {author} {\bibinfo {author} {\bibfnamefont {B.}~\bibnamefont
  {Sutherland}},\ }\href {\doibase 10.1103/PhysRevA.4.2019} {\bibfield
  {journal} {\bibinfo  {journal} {Phys. Rev. A}\ }\textbf {\bibinfo {volume}
  {4}},\ \bibinfo {pages} {2019} (\bibinfo {year} {1971})}\BibitemShut
  {NoStop}%
\bibitem [{\citenamefont {Haldane}(1988)}]{Haldane1988}%
  \BibitemOpen
  \bibfield  {author} {\bibinfo {author} {\bibfnamefont {F.~D.~M.}\
  \bibnamefont {Haldane}},\ }\href {\doibase 10.1103/PhysRevLett.60.635}
  {\bibfield  {journal} {\bibinfo  {journal} {Phys. Rev. Lett.}\ }\textbf
  {\bibinfo {volume} {60}},\ \bibinfo {pages} {635} (\bibinfo {year}
  {1988})}\BibitemShut {NoStop}%
\bibitem [{\citenamefont {Shastry}(1988)}]{Shastry1988}%
  \BibitemOpen
  \bibfield  {author} {\bibinfo {author} {\bibfnamefont {B.~S.}\ \bibnamefont
  {Shastry}},\ }\href {\doibase 10.1103/PhysRevLett.60.639} {\bibfield
  {journal} {\bibinfo  {journal} {Phys. Rev. Lett.}\ }\textbf {\bibinfo
  {volume} {60}},\ \bibinfo {pages} {639} (\bibinfo {year} {1988})}\BibitemShut
  {NoStop}%
\bibitem [{\citenamefont {Giamarchi}(2003)}]{Giamarchi-book}%
  \BibitemOpen
  \bibfield  {author} {\bibinfo {author} {\bibfnamefont {T.}~\bibnamefont
  {Giamarchi}},\ }\href@noop {} {\emph {\bibinfo {title} {Quantum Physics in
  One Dimension}}}\ (\bibinfo  {publisher} {Oxford University Press},\ \bibinfo
  {address} {Oxford},\ \bibinfo {year} {2003})\BibitemShut {NoStop}%
\bibitem [{\citenamefont {Calabrese}\ and\ \citenamefont
  {Cardy}(2004)}]{CalabreseC2004}%
  \BibitemOpen
  \bibfield  {author} {\bibinfo {author} {\bibfnamefont {P.}~\bibnamefont
  {Calabrese}}\ and\ \bibinfo {author} {\bibfnamefont {J.}~\bibnamefont
  {Cardy}},\ }\href {\doibase 10.1088/1742-5468/2004/06/p06002} {\bibfield
  {journal} {\bibinfo  {journal} {Journal of Statistical Mechanics: Theory and
  Experiment}\ }\textbf {\bibinfo {volume} {2004}},\ \bibinfo {pages} {P06002}
  (\bibinfo {year} {2004})}\BibitemShut {NoStop}%
\bibitem [{\citenamefont {Capello}\ \emph {et~al.}(2005)\citenamefont
  {Capello}, \citenamefont {Becca}, \citenamefont {Yunoki}, \citenamefont
  {Fabrizio},\ and\ \citenamefont {Sorella}}]{Capelloetal2005}%
  \BibitemOpen
  \bibfield  {author} {\bibinfo {author} {\bibfnamefont {M.}~\bibnamefont
  {Capello}}, \bibinfo {author} {\bibfnamefont {F.}~\bibnamefont {Becca}},
  \bibinfo {author} {\bibfnamefont {S.}~\bibnamefont {Yunoki}}, \bibinfo
  {author} {\bibfnamefont {M.}~\bibnamefont {Fabrizio}}, \ and\ \bibinfo
  {author} {\bibfnamefont {S.}~\bibnamefont {Sorella}},\ }\href {\doibase
  10.1103/PhysRevB.72.085121} {\bibfield  {journal} {\bibinfo  {journal} {Phys.
  Rev. B}\ }\textbf {\bibinfo {volume} {72}},\ \bibinfo {pages} {085121}
  (\bibinfo {year} {2005})}\BibitemShut {NoStop}%
\bibitem [{\citenamefont {Capello}\ \emph {et~al.}(2008)\citenamefont
  {Capello}, \citenamefont {Becca}, \citenamefont {Fabrizio},\ and\
  \citenamefont {Sorella}}]{Capelloetal2008}%
  \BibitemOpen
  \bibfield  {author} {\bibinfo {author} {\bibfnamefont {M.}~\bibnamefont
  {Capello}}, \bibinfo {author} {\bibfnamefont {F.}~\bibnamefont {Becca}},
  \bibinfo {author} {\bibfnamefont {M.}~\bibnamefont {Fabrizio}}, \ and\
  \bibinfo {author} {\bibfnamefont {S.}~\bibnamefont {Sorella}},\ }\href
  {\doibase 10.1103/PhysRevB.77.144517} {\bibfield  {journal} {\bibinfo
  {journal} {Phys. Rev. B}\ }\textbf {\bibinfo {volume} {77}},\ \bibinfo
  {pages} {144517} (\bibinfo {year} {2008})}\BibitemShut {NoStop}%
\bibitem [{\citenamefont {Eggert}(1996)}]{Eggert1996}%
  \BibitemOpen
  \bibfield  {author} {\bibinfo {author} {\bibfnamefont {S.}~\bibnamefont
  {Eggert}},\ }\href {\doibase 10.1103/PhysRevB.54.R9612} {\bibfield  {journal}
  {\bibinfo  {journal} {Phys. Rev. B}\ }\textbf {\bibinfo {volume} {54}},\
  \bibinfo {pages} {R9612} (\bibinfo {year} {1996})}\BibitemShut {NoStop}%
\bibitem [{\citenamefont {Majumdar}\ and\ \citenamefont
  {Ghosh}(1969)}]{MajumdarG1969}%
  \BibitemOpen
  \bibfield  {author} {\bibinfo {author} {\bibfnamefont {C.~K.}\ \bibnamefont
  {Majumdar}}\ and\ \bibinfo {author} {\bibfnamefont {D.~K.}\ \bibnamefont
  {Ghosh}},\ }\href {\doibase 10.1063/1.1664978} {\bibfield  {journal}
  {\bibinfo  {journal} {Journal of Mathematical Physics}\ }\textbf {\bibinfo
  {volume} {10}},\ \bibinfo {pages} {1388} (\bibinfo {year} {1969})},\ \Eprint
  {http://arxiv.org/abs/https://doi.org/10.1063/1.1664978}
  {https://doi.org/10.1063/1.1664978} \BibitemShut {NoStop}%
\bibitem [{\citenamefont {Bursill}\ \emph {et~al.}(1995)\citenamefont
  {Bursill}, \citenamefont {Gehring}, \citenamefont {Farnell}, \citenamefont
  {Parkinson}, \citenamefont {Xiang},\ and\ \citenamefont
  {Zeng}}]{Bursilletal1995}%
  \BibitemOpen
  \bibfield  {author} {\bibinfo {author} {\bibfnamefont {R.}~\bibnamefont
  {Bursill}}, \bibinfo {author} {\bibfnamefont {G.~A.}\ \bibnamefont
  {Gehring}}, \bibinfo {author} {\bibfnamefont {D.~J.~J.}\ \bibnamefont
  {Farnell}}, \bibinfo {author} {\bibfnamefont {J.~B.}\ \bibnamefont
  {Parkinson}}, \bibinfo {author} {\bibfnamefont {T.}~\bibnamefont {Xiang}}, \
  and\ \bibinfo {author} {\bibfnamefont {C.}~\bibnamefont {Zeng}},\ }\href
  {\doibase 10.1088/0953-8984/7/45/016} {\bibfield  {journal} {\bibinfo
  {journal} {Journal of Physics: Condensed Matter}\ }\textbf {\bibinfo {volume}
  {7}},\ \bibinfo {pages} {8605} (\bibinfo {year} {1995})}\BibitemShut
  {NoStop}%
\bibitem [{\citenamefont {White}\ and\ \citenamefont
  {Affleck}(1996)}]{WhiteA1996}%
  \BibitemOpen
  \bibfield  {author} {\bibinfo {author} {\bibfnamefont {S.~R.}\ \bibnamefont
  {White}}\ and\ \bibinfo {author} {\bibfnamefont {I.}~\bibnamefont
  {Affleck}},\ }\href {\doibase 10.1103/PhysRevB.54.9862} {\bibfield  {journal}
  {\bibinfo  {journal} {Phys. Rev. B}\ }\textbf {\bibinfo {volume} {54}},\
  \bibinfo {pages} {9862} (\bibinfo {year} {1996})}\BibitemShut {NoStop}%
\bibitem [{\citenamefont {Gehring}\ \emph {et~al.}(1997)\citenamefont
  {Gehring}, \citenamefont {Bursill},\ and\ \citenamefont
  {Xiang}}]{Gehringetal1997}%
  \BibitemOpen
  \bibfield  {author} {\bibinfo {author} {\bibfnamefont {G.~A.}\ \bibnamefont
  {Gehring}}, \bibinfo {author} {\bibfnamefont {R.~J.}\ \bibnamefont
  {Bursill}}, \ and\ \bibinfo {author} {\bibfnamefont {T.}~\bibnamefont
  {Xiang}},\ }\href@noop {} {\bibfield  {journal} {\bibinfo  {journal} {Acta
  Physica Polonica A}\ }\textbf {\bibinfo {volume} {91}},\ \bibinfo {pages}
  {105} (\bibinfo {year} {1997})}\BibitemShut {NoStop}%
\bibitem [{\citenamefont {Deschner}\ and\ \citenamefont
  {S\o{}rensen}(2013)}]{DeschnerS2013}%
  \BibitemOpen
  \bibfield  {author} {\bibinfo {author} {\bibfnamefont {A.}~\bibnamefont
  {Deschner}}\ and\ \bibinfo {author} {\bibfnamefont {E.~S.}\ \bibnamefont
  {S\o{}rensen}},\ }\href {\doibase 10.1103/PhysRevB.87.094415} {\bibfield
  {journal} {\bibinfo  {journal} {Phys. Rev. B}\ }\textbf {\bibinfo {volume}
  {87}},\ \bibinfo {pages} {094415} (\bibinfo {year} {2013})}\BibitemShut
  {NoStop}%
\bibitem [{\citenamefont {Ferrari}\ \emph {et~al.}(2018)\citenamefont
  {Ferrari}, \citenamefont {Parola}, \citenamefont {Sorella},\ and\
  \citenamefont {Becca}}]{Ferrarietal2018}%
  \BibitemOpen
  \bibfield  {author} {\bibinfo {author} {\bibfnamefont {F.}~\bibnamefont
  {Ferrari}}, \bibinfo {author} {\bibfnamefont {A.}~\bibnamefont {Parola}},
  \bibinfo {author} {\bibfnamefont {S.}~\bibnamefont {Sorella}}, \ and\
  \bibinfo {author} {\bibfnamefont {F.}~\bibnamefont {Becca}},\ }\href
  {\doibase 10.1103/PhysRevB.97.235103} {\bibfield  {journal} {\bibinfo
  {journal} {Phys. Rev. B}\ }\textbf {\bibinfo {volume} {97}},\ \bibinfo
  {pages} {235103} (\bibinfo {year} {2018})}\BibitemShut {NoStop}%
\bibitem [{\citenamefont {Zeng}\ and\ \citenamefont
  {Parkinson}(1995)}]{ZengP1995}%
  \BibitemOpen
  \bibfield  {author} {\bibinfo {author} {\bibfnamefont {C.}~\bibnamefont
  {Zeng}}\ and\ \bibinfo {author} {\bibfnamefont {J.~B.}\ \bibnamefont
  {Parkinson}},\ }\href {\doibase 10.1103/PhysRevB.51.11609} {\bibfield
  {journal} {\bibinfo  {journal} {Phys. Rev. B}\ }\textbf {\bibinfo {volume}
  {51}},\ \bibinfo {pages} {11609} (\bibinfo {year} {1995})}\BibitemShut
  {NoStop}%
\bibitem [{\citenamefont {Vodola}()}]{Vodola}%
  \BibitemOpen
  \bibfield  {author} {\bibinfo {author} {\bibfnamefont {D.}~\bibnamefont
  {Vodola}},\ }\href@noop {} {}\bibinfo {howpublished} {personal
  communication}\BibitemShut {NoStop}%
\bibitem [{\citenamefont {Cazalilla}(2004)}]{Cazalilla2004}%
  \BibitemOpen
  \bibfield  {author} {\bibinfo {author} {\bibfnamefont {M.~A.}\ \bibnamefont
  {Cazalilla}},\ }\href {\doibase 10.1088/0953-4075/37/7/051} {\bibfield
  {journal} {\bibinfo  {journal} {Journal of Physics B: Atomic, Molecular and
  Optical Physics}\ }\textbf {\bibinfo {volume} {37}},\ \bibinfo {pages} {S1}
  (\bibinfo {year} {2004})}\BibitemShut {NoStop}%
\bibitem [{\citenamefont {L\"auchli}\ and\ \citenamefont
  {Kollath}(2008)}]{LauchliK2008}%
  \BibitemOpen
  \bibfield  {author} {\bibinfo {author} {\bibfnamefont {A.~M.}\ \bibnamefont
  {L\"auchli}}\ and\ \bibinfo {author} {\bibfnamefont {C.}~\bibnamefont
  {Kollath}},\ }\href {\doibase 10.1088/1742-5468/2008/05/p05018} {\bibfield
  {journal} {\bibinfo  {journal} {Journal of Statistical Mechanics: Theory and
  Experiment}\ }\textbf {\bibinfo {volume} {2008}},\ \bibinfo {pages} {P05018}
  (\bibinfo {year} {2008})}\BibitemShut {NoStop}%
\end{thebibliography}%

\end{document}